\documentclass{ieeeaccess}
\usepackage{amsmath,amssymb,amsfonts,amsthm,mathrsfs}
\usepackage{algorithmic}
\usepackage{graphicx}
\usepackage{textcomp}
\usepackage{multirow}
\usepackage[caption=false, font=footnotesize]{subfig}
\usepackage{algorithm}
\usepackage{algorithmic}
\usepackage{bm}
\usepackage{float}
\usepackage{array,tabularx}
\usepackage{url}
\usepackage{soul}
\usepackage{wasysym}

\usepackage[backend=bibtex,sorting=none]{biblatex}
\newcommand{\fref}[1]{Fig.~\ref{#1}}
\newcommand{\bx}{\bm{x}}
\newcommand{\change}[1]{{#1}}
\newcommand{\HJadd}[1]{{#1}}
\newcommand{\JKadd}[1]{{#1}}

\makeatletter
\AtBeginDocument{\DeclareMathVersion{bold}
\SetSymbolFont{operators}{bold}{T1}{times}{b}{n}
\SetSymbolFont{NewLetters}{bold}{T1}{times}{b}{it}
\SetMathAlphabet{\mathrm}{bold}{T1}{times}{b}{n}
\SetMathAlphabet{\mathit}{bold}{T1}{times}{b}{it}
\SetMathAlphabet{\mathbf}{bold}{T1}{times}{b}{n}
\SetMathAlphabet{\mathtt}{bold}{OT1}{pcr}{b}{n}
\SetSymbolFont{symbols}{bold}{OMS}{cmsy}{b}{n}
\renewcommand\boldmath{\@nomath\boldmath\mathversion{bold}}}
\makeatother

\def\BibTeX{{\rm B\kern-.05em{\sc i\kern-.025em b}\kern-.08em
    T\kern-.1667em\lower.7ex\hbox{E}\kern-.125emX}}

\begin{document}
\history{Received 12 July 2024, accepted 12 August 2024, date of publication 15 August 2024, date of current version 11 September 2024}
\doi{10.1109/ACCESS.2023.1120000}

\title{Fast Marching based Rendezvous Path Planning for a Team of Heterogeneous Vehicles}
\author{\uppercase{Jaekwang Kim}\authorrefmark{1},
\uppercase{Hyung-Jun Park}\authorrefmark{2}, \uppercase{Aditya Penumarti}\authorrefmark{3}, and \uppercase{Jaejeong (Jane) Shin}\authorrefmark{4},
\IEEEmembership{Member, IEEE}}

\address[1]{Department of Mechanical and Design Engineering, Hongik University, Sejong 30016 Republic of Korea (e-mail: jk12@hongik.ac.kr)}
\address[2]{School of Mechanical and Aerospace Engineering, Sunchon National University, Sunchon Junnam 57922 Republic of Korea (e-mail: hjpark89@scnu.ac.kr)}
\address[3]{Department of Mechanical and Aerospace Engineering, University of Florida, Gainesville FL 32611 USA (e-mail: apenumarti@ufl.edu)}
\address[4]{Department of Mechanical and Aerospace Engineering, University of Florida, Gainesville FL 32611 USA (e-mail: jane.shin@ufl.edu)}

\tfootnote{This work was partially supported by the Hongik University New Faculty 
Research Support Fund and the National Research Foundation of Korea Grant 
funded by the Korean Government (No. RS-2024-00333943)}

\markboth
{J. Kim \headeretal: Fast Marching based Rendezvous Path Planning for a Team of Heterogeneous Vehicles}
{J. Kim \headeretal: Fast Marching based Rendezvous Path Planning for a Team of Heterogeneous Vehicles}

\corresp{Corresponding author: Jaejeong (Jane) Shin (e-mail: jane.shin@ufl.edu).}

\begin{abstract}
\change{This paper presents a} formulation\change{ }for deterministically calculating\change{ }optimized paths for a multi-agent system consisting of heterogeneous vehicles. The key idea is the calculation of the shortest time for each agent to reach every grid point from its known initial position. \change{Such} arrival time map \change{is efficiently computed} using the Fast Marching Method (FMM), a computational algorithm originally designed for solving boundary value problems of the Eikonal equation. \change{By l}everaging the FMM\change{,} we demonstrate that the minimal time rendezvous point and paths for all member vehicles can be uniquely determined with minimal computational \change{overhead}. 
\change{
The scalability and adaptability of the present method during online execution are investigated, followed by a comparison with a baseline method that highlights the effectiveness of the proposed approach.}
Then, the potential of the present method is showcased \change{through} a virtual rendezvous scenario \change{involving} the coordination of a ship, an underwater vehicle, an aerial vehicle, and a ground vehicle, \change{all converging} at the optimal location within the Tampa Bay area in minimal time. The results show that the developed framework can efficiently construct continuous paths of heterogeneous vehicles by accommodating operational constraints via an FMM algorithm.

\end{abstract}


\titlepgskip=-21pt

\maketitle

\section{Introduction}
\label{sec:introduction}
Recent advancements in various types of autonomous vehicles have sparked interest in multi-agent systems, which hold the potential to efficiently address complex tasks. Strategic multi-agent path finding (MAPF) becomes crucial, particularly when the team comprises heterogeneous vehicles with varying operational domains and capabilities, such as different speeds, sizes, and maneuverability. These agents may encompass a wide range of vehicles, including ships, underwater vehicles, aerial vehicles, and ground vehicles. Each type of vehicle can have unique navigational \change{and environmental} constraints \change{depending on each one's operation domain} \cite{ShinIMVP22, DiazBathydrone22}. \change{Incorporating heterogeneous vehicles across multiple domains enhances the system's ability to handle complex and large-scale operations, significantly impacting real-world applications such as autonomous vehicle fleets for delivery, disaster response teams, and environmental monitoring.} 

Previous studies in multi-agent planning \change{for homogeneous systems} have primarily concentrated on scheduling, with an emphasis on task allocation and agent coordination~\change{\mbox{\cite{thompson2019review, baysPersistentScheduleEvaluation2024}}}. \change{As scheduling algorithms focus on finding the optimal sequences and coordination among heterogeneous vehicles, the continuous path planning of each agent is often neglected and considered as a lower level problem. However, considering continuous path planning}\footnote{Here, a continuous path refers to a path defined on continuous real-world space and thus can serve as a smooth path for autonomous vehicles.} ensures smooth and uninterrupted motion of vehicles~\cite{MelchiorFMCompStudyObstacleDanger03}, since all agents in real-world applications must adapt their paths in response to changing environmental conditions and dynamic obstacles. \change{Similar problems have been considered as rendezvous (RDV) search problems as well. However, rendezvous search approaches in existing literature focus on finding the optimal strategy for rendezvous given limited sensing and communication capabilities among homogeneous agents \mbox{\cite{lin2003multi, fang2008multi, lin2005multi, ozsoyeller2022m, anderson1990rendezvous, pelc2019deterministic, ta2014deterministic, dereniowski2015rendezvous}}. The common assumptions in rendezvous search approaches, such as unknown environment, unknown initial conditions, and asynchronous systems, are important to consider in certain scenarios; however, these assumptions may be relaxed in a larger-scale scenarios where heterogeneous systems are usually deployed.} 

\change{The aforementioned rendezvous path planning can also be viewed from} the perspective of computational science, \change{as} a general form of \change{the continuous MAPF problem, which} is known as an NP-hard problem~\cite{MAPF1}. \change{The MAPF problems, however, mostly focus on collision avoidance among the agents. Therefore, it is not possible to directly apply existing MAPF methods to the considered rendezvous path planning problem. Moreover, o}ne of the main challenges in MAPF is related to the high dimensionality of the problem. With many agents in an environment, the number of potential paths and interactions can become overwhelmingly large. The complexity of the multi-agent path finding problem also stiffly increases, as the problem as the number of agents in the system increases, and thus solving MAPF problems becomes computationally expensive. \JKadd{In addition}, in many cases, efficiency is not the sole concern; safety (collision-free paths) must also be taken into account. Due to the complex nature and conflicting objectives encountered in MAPF problems, one \JKadd{often} needs to reduce or approximate the original problem to a simpler form, compromising accuracy and global optimality. \change{The comparison between the existing and presented problems is summarized in Table \mbox{\ref{tab:comparison}}. The table shows that why the existing approaches cannot be directly applied to solve the rendezvous path planning problem considered in this paper.} 

In this work, we consider continuous path planning \change{for} a multi-agent system for minimal time rendezvous tasks. In these tasks, some agents initially operating at different locations are tasked with meeting to exchange information or resources. Such\change{ }tasks are frequently encountered in  spacecraft docking scenarios~\cite{Rendez1,Rendez2}. A team of autonomous underwater vehicles also often needs to initiate information exchange tasks at close distances due to limited data transfer capabilities in deep water~\cite{Rendez3}. In these scenarios, identifying the optimal rendezvous point and the path for each agent to achieve the earliest possible rendezvous time as a team (or other optimizing goals) is important. Unfortunately, however, planning paths that accommodate differences of vehicles, while optimizing overall performance remains a significant challenge. 


The primary contribution of this paper lies in formulating the rendezvous problem \change{for} a multi-agent system in a way that is suitable for assessment using the fast marching method (FMM). The FMM is a well-established numerical technique originally developed for solving the Eikonal equation. Beyond its original purpose, however, the FMM has also demonstrated its capability in efficiently computing the shortest paths on continuous grids~\cite{SongTVMarineEnv17, LiuPredictiveTimeVaryingEnv17, YanAnisotropicFMSingle20, ZhangAnisotropicFMBridge23, ChenFMSandVelocityObs20}. Extending these works, we show how the use of the FMM for rendezvous MAPF also enables the enhancement of collaboration, reduction of complexity, and optimization of the overall mission performance of the team. Specifically, we first define an optimization problem that involves continuous path planning for a team of heterogeneous vehicles, each with its \change{own} operational domain. Then, we exploit the direct output from the FMM as a key component of a new path planning approach. Our approach deterministically calculates the time-optimal rendezvous point for heterogeneous vehicles and determines the path to the optimal rendezvous point from different initial agent positions. Throughout this process, the method also takes into account their unique operational constraints. 

The remainder of the paper is organized as follows. Section~\ref{sec:background} introduces \change{the} methodologies of the FMM and FMM-based path planning. In Section~\ref{sec:method}, we formulate an optimization problem for multi-agent path planning of a rendezvous task and introduce a new methodology to efficiently solve the problem. Section~\ref{sec:experiment} presents a virtual path planning experiment to demonstrate the potential of our proposed approach, while Section~\ref{sec:discussion} discusses important features and highlights \change{the} merits of the suggested methodology. Finally, we conclude the paper in Section~\ref{sec:conclusion}, listing potential future research directions. 

\begin{table}[]
    \centering
    \begin{tabular}{|c||c|c|c|c|}
        \hline
        Approaches & RDV & Path Finding & 
        Heterogeneous & Obs. \\
        \hline
        \hline
        Scheduling & $\ocircle$ & $\times$ & $\ocircle$ & $\times$\\
        \hline
        RDV Search & $\ocircle$ & $\times$ & $\times$ & $\ocircle$ \\
        \hline
        MAPF & $\times$ & $\ocircle$ & $\times$ & $\ocircle$ \\
        \hline
        Presented & $\ocircle$ & $\ocircle$ & $\ocircle$ & $\ocircle$ \\
        \hline
    \end{tabular}
    \caption{\change{Comparison of existing approaches to the presented FMM-based method on the rendezvous path planning problem for heterogeneous vehicle systems. The first column (RDV) indicates whether the approach can consider rendezvous point search. The second column (Path Finding) shows if the approach can compute a continuous trajectory in the environment. The third column (Heterogeneous) specifies whether the approach can apply to a multi-agent system consisting of heterogeneous agents operating in multiple domains. The fourth column (Obs.) denotes if the method can consider obstacles in the environment.}}
    \label{tab:comparison}
    \vspace*{-0.7cm}
\end{table}

\section{Background on Fast Marching Method and Its Application to Path Optimization}
\label{sec:background}

In this section, we provide a brief overview of the FMM, which will be used to address the challenges of multi-agent path planning for rendezvous missions. Originally developed for solving a nonlinear first-order partial differential equation, the FMM has shown high efficiency in dealing with interface mechanics compared to other algorithms~\cite{Park:2021,Park:2023, DiffuseInterface,PESKIN1972252}. The FMM has also found applications in diverse research domains, encompassing materials science~\cite{JKim:2021}, computer graphics~\cite{FMMgraphics}, \HJadd{wave propagation~\mbox{\cite{jiang2021fast}}}, and image processing~\cite{FMMimage,forcadel2008generalized}. Particularly, its application to path optimization has a long history in various domains of applications, ranging from marine vehicles to social navigation ~\cite{Gomez,Garrido:2006,Garrido:2009,Garrido:2013,Garrido:2015,gomez2014fast}. \HJadd{Since the FMM is a numerical technique for solving the propagation of interfaces (or waves), its applications are determined by how the speed function for interface propagation is defined. For instance, in the field of image segmentation, the sign of the speed function flips when the interface reaches an object, causing the interface to propagate outward along the object's boundary. On the other hand, in path planning, a speed function based on the distance to obstacles is often used to dampen the speed of a vehicle as it approaches an obstacle, as detailed in Section \mbox{\ref{ssec:adaptation}}. While the FMM can be applied across various research fields, the assumption that the medium through which interfaces propagate is homogeneous and isotropic remains a challenging issue for specific applications.} In the following, we begin by summarizing the main ideas of the FMM in its original context. 

\subsection{The fast marching method}
\begin{figure}
    \centering
    \includegraphics[width=0.6\linewidth]{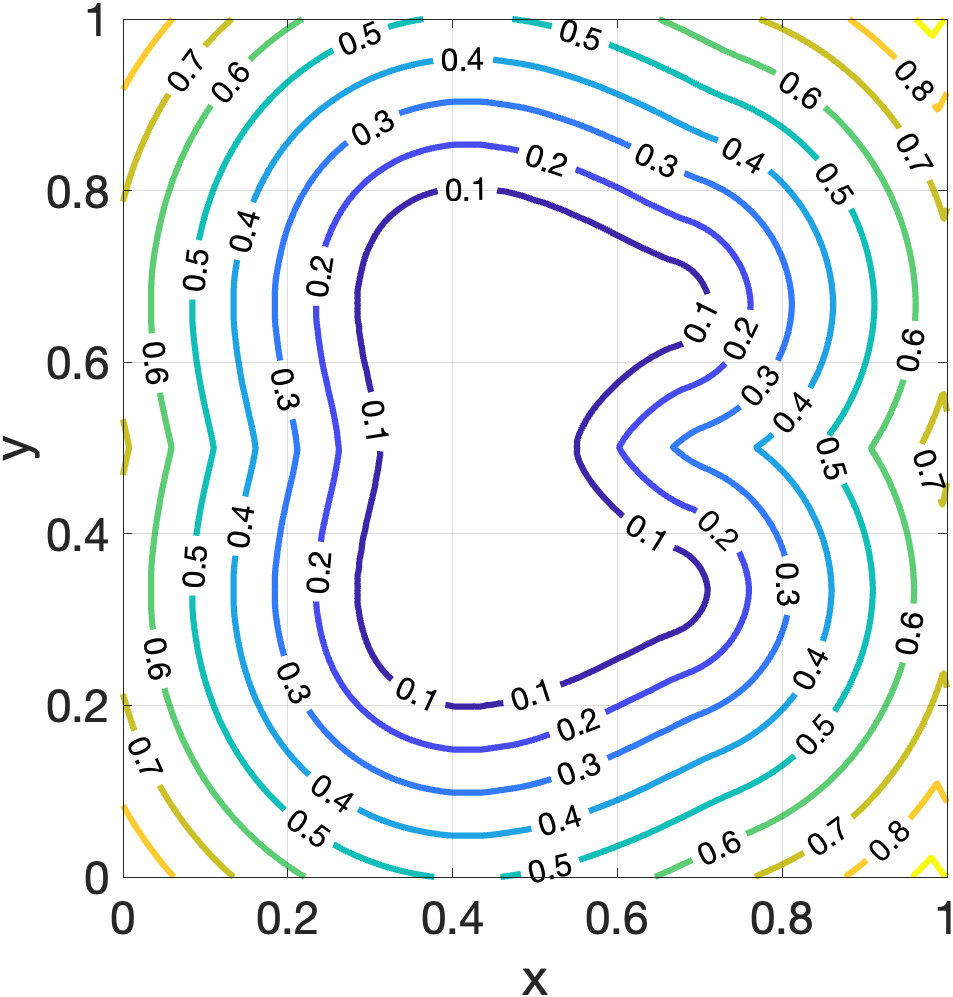}
  \caption{The level sets of the solution to the Eikonal equation~\eqref{e:eikonal} computed using the fast marching method, describe a surface evolving with outward normal velocity $V(\bx)$ = 1. The level set values are indicative of the time it takes for the initial surface (represented by the innermost blue line) to reach each grid point within the computational domain.}
  \label{fig:fmm_surface}
\end{figure}

First introduced in\change{~}\cite{Tsitsiklisl}, the FMM is an efficient computational algorithm for tracking the front, or \textit{interface}, that evolves with the outward unit normal direction with speed $V$. The explicit outcome of FMM is the arrival time $T(\bx)$ that the initial surface needs to reach every point $\bx$ \change{in} the given domain $\Omega$. For example, \fref{fig:fmm_surface} demonstrates the result of FMM used to track an initial surface $\Gamma$ (the innermost blue line) growing with a uniform outward normal velocity $V(\bx)=1$. Below, we \change{outline} the FMM algorithm as described in\change{ }\cite{Sethian}. 

Let $s(t)$ describe a surface evolving \change{at} speed $V(\bx)$ from a given initial surface $s(0)=\Gamma$. Instead of solving a time-dependent problem for $s(t)$ to track the moving surface, the FMM solves a function $T(\bx)$ defined as
\begin{equation}
    T(s(t)) = t,
    \label{e:zeta_def}
\end{equation}
with $T = 0 $ on $\Gamma$. Differentiating \eqref{e:zeta_def} and noting that $\nabla T$ is normal to the surface, one arrives at the following boundary value problem,
\begin{equation}
    |\nabla T|V=1. 
    \label{e:eikonal}
\end{equation}
Also, the boundary condition for $T$ equivalent to the original time-dependent problem is 
\begin{equation}
    T=0 \quad \text{on } \Gamma. 
\end{equation}
Equation \eqref{e:eikonal} is commonly referred to as the Eikonal equation. 

Now, we describe the algorithm to solve \eqref{e:eikonal} on a two-dimensional discrete grid, i.e.\change{,} $\bx = (x,y)$. However, it is worth noting that the algorithm \change{can be} conveniently generalized to arbitrary dimensions. Let $D^{-x}_{ij} (\cdot)$ denote\change{ }the standard backward-difference operation on the grid point $ij$
\begin{equation}
    D^{-x}_{ij}T = \frac{T_{ij}-T_{(i-1)j}}{\delta x}.
\end{equation}
Likewise, we use $D^{+x}, D^{-y}$, and $D^{+y}$ to represent forward in $x$, backward and forward in $y$ backward finite difference operators, respectively. In order to ensure a unique \textit{viscosity solution} for the Eikonal equation~\eqref{e:eikonal}, we necessitate the consistent utilization of an upwind finite difference scheme when computing the gradient. This \change{step} is compactly written as
\begin{equation}
\begin{split}
    \frac{1}{V(\bx)} = 
    \big[(\mathrm{max} (D^{-x}_{ij}T,-D^{+x}T_{ij}, 0)^2  \\
    + \mathrm{max} (D^{-y}_{ij}T,-D^{+y}T_{ij}, 0)^2 \big] ^{1/2}
    \label{e:eikonal_algebraic}
\end{split}
\end{equation}
When the neighboring values of $T_{ij}$ are known, the discrete Eikonal equation~\eqref{e:eikonal_algebraic} becomes a quadratic equation for $T_{ij}$ at each grid point, allowing for straightforward analytical solutions.

The FMM initiates by performing the following initialization step.
\begin{enumerate}
    \item Assign $T (\bx) =0 $ for grid points in the area enclosed by the initial surface, and tag them as \textit{accepted}\change{.}
    \item Assign $T (\bx) = + \infty$ for the remaining grid points, and tag them as \textit{far}\change{.}
    \item Among the \textit{accepted} points, identify the points that are in the neighborhood of points tagged as \textit{far}, and tag them as \textit{considered}\change{.}
\end{enumerate}

The key step in the fast marching method is to update $T$ with a trial value using Eq. \eqref{e:eikonal_algebraic} for grid points tagged\HJadd{ }as \textit{considered}, while only accepting the update with the smallest value at each iteration. This procedure requires keeping track of the smallest $T$-value among points tagged as \textit{considered}. The potential $T$ values are managed in a specialized data structure inspired by discrete network algorithms~\cite{Heap}. This data structure is known as a min-heap data structure, which represents a complete binary tree with a property that the value at any given node is less than or equal to the values of its children. Utilizing the min-heap, the FMM then proceeds as follows.
\begin{enumerate}
    \item Form a min-heap structure for the \textit{considered} points.
    \item Access the minimum value of the heap, located at the root of the binary tree. 
    \item Determine a trial solution $\tilde{T}$ on the neighbors of the root using \eqref{e:eikonal_algebraic}.
            If the trial solution $\tilde{T}$ is smaller than the present values, then update $T(\bx)=\tilde{T}$.
    \item If a point, previously tagged as \textit{far}, is updated using a trial value, 
         relabel it as \textit{considered}, and add it to the heap structure.
    \item Tag the root of the heap as \textit{accepted}, and delete it from the heap.
    \item Repeat steps 2 to 5, until every grid point is tagged as \textit{accepted}.
\end{enumerate}

\HJadd{In a min-heap, the time complexity for insertion and deletion operations is $\mathcal{O}(\log {{n}^H_{e}})$, where ${{n}^H_{e}}$ is the number of elements in the heap, and $\mathcal{O}(\cdot )$ denotes Big O notation, a limiting behavior of a function. The notation expresses an upper bound on the execution time required by an algorithm. The time complexity of the mean-heap arises from its binary tree structure, which ensures that the height of the tree is $\log {{n}^H_{e}}$. Consequently, only a logarithmic number of comparisons are needed for adding or removing elements, enhancing the efficiency of these operations. In the FMM, every grid point has a mean-heap structure, resulting in a time complexity of $\mathcal{O}(n\log n)$, where $n$ is the total number of grid points.} 

\subsection{Adaptation of the FMM for path optimization}
\label{ssec:adaptation}
While the FMM is originally developed for interface problems, numerous studies have also successfully applied the FMM in vehicle path planning scenarios, enabling agents to navigate complex environments, avoid obstacles, and reach their destinations efficiently. These studies have primarily focused on single-agent path planning under various external conditions. These include time-varying environmental factors, such as waves and currents in oceans~\cite{SongTVMarineEnv17}, time-varying environments with predictive models~\cite{LiuPredictiveTimeVaryingEnv17}, angle guidance for uncrewed surface vehicles~\cite{LiuUSVHeadingAngle16}, \change{the} anisotropic Fast Marching (FM)-based approaches for dynamic obstacles~\cite{YanAnisotropicFMSingle20} and bridge obstacles~\cite{ZhangAnisotropicFMBridge23}, as well as path planning for autonomous ships~\cite{ChenFMSandVelocityObs20}. In contrast, its application in multi-agent systems remains relatively unexplored. A few examples include swarm coordination~\cite{TanSwarmUSV20} and formation control involving vehicles with different dynamic properties~\cite{TanHeterogeneousUSV23}.

\begin{figure}
    \centering
    \includegraphics[width=0.5\linewidth]{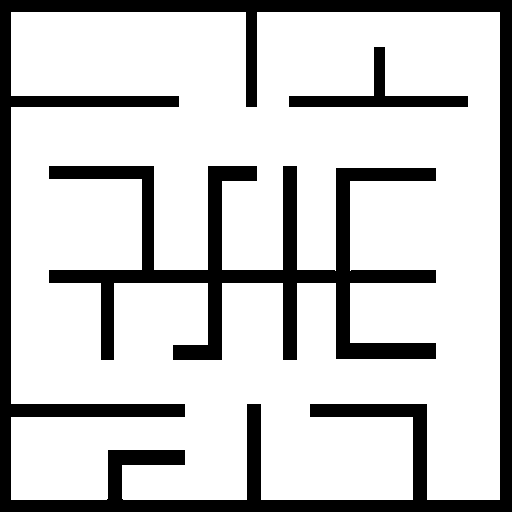}
    \caption{An example of binary occupancy map. The binary image, which is on $512 \times 512$ pixel size, takes the value of 0 if the position is occupied by obstacles, and 1 otherwise.}
    \label{fig:binaryOccupancymap}
\end{figure}

In the context of path optimization, the computational domain $\Omega$ of the FMM takes on a new perspective as the configuration space for mobile agents, often depicted through a binary occupancy map as illustrated in \fref{fig:binaryOccupancymap}. The binary image, which is in a size of $n=n_1\times n_2$ pixels, takes the value of 0 if the position is occupied by obstacles, and 1 otherwise. Also, the initial surface $\Gamma$ is reduced to a single wave-source point $\bx_0$, representing the initial location of an agent. The velocity field $V(\bx)$ signifies the permissible speed of vehicles at a given position while considering the proximity of obstacles (such as walls and barriers) to the agents. As part of the FMM's initialization step, every grid point located on obstacles is initially labeled as\change{ }\textit{accepted}\change{.} 

Next, the FMM algorithm is executed to compute the shortest time $T(\bx)$ for the propagating wave to arrive at each grid point. The trajectory of the agent is finally determined by extracting the maximum gradient direction of $T(\bx)$ from the target point to the initial point. Since $T(\bx)$ is derived from the target point, the resulting $T$-field uniquely exhibits a single minimum at the target point, ensuring a unique solution~\cite{Garrido:2009}. 

A remaining task is to employ an appropriate model for the velocity field $V(\bx)$ that respects the environment. While one might simply consider the simplest option, which is to use a constant value $\mathcal{V}_{max}$ representing the maximum speed of the agents, it is observed that the resulting trajectory lacks realism as it fails to ensure both smoothness and a safe distance between agents and obstacles~\cite{Gomez}. To address these issues, the FMM has been advanced into the Fast Marching Square (FMS) method. In order to guarantee a safe distance between obstacles and agents, this approach introduces a penalty to the agent's velocity as it navigates in proximity to obstacles. The FMS method entails the implementation of two distinct FMMs. 

The objective of the first FMM is to construct a velocity grid map that takes into account the presence of obstacles. This \change{objective} is achieved by evolving initial surfaces, which represent the boundaries of obstacles in the environment, with a constant velocity $V(\bx) = 1$. The outcome of this process is the computation of the distance $d(\bx) = T(\bx) \in \mathbb R^+$ at each grid point, indicating the shortest distance to the nearest obstacle. Consequently, a velocity grid map $V$ is computed as a function of $d$, which is artificially designed to penalize the vehicle's speed as it approaches obstacles. A common choice for this penalizing function is a linear relationship $V \propto d$, which is inspired by a two-dimensional repulsive electrostatic potential~\cite{Garrido:2009_2}. 

\begin{figure}
    \centering
    \includegraphics[width=0.8\linewidth]{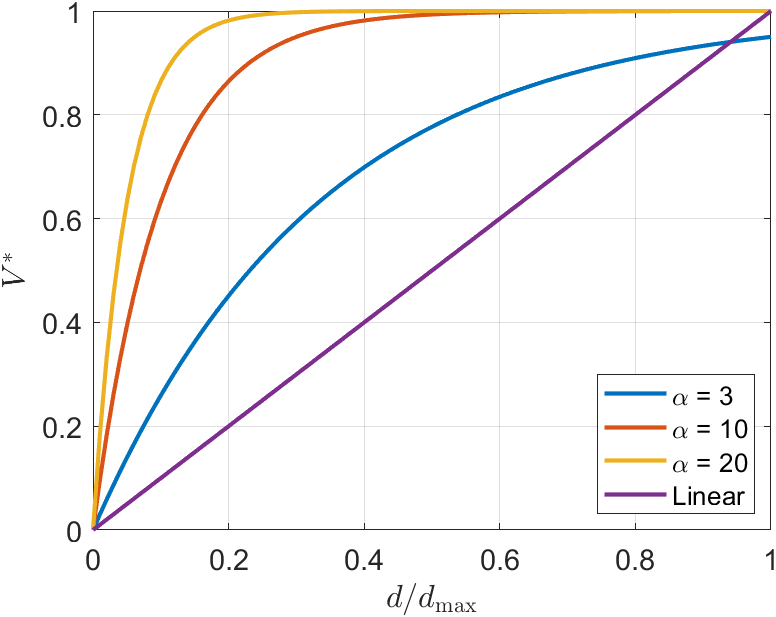}
    \caption{Plots of velocity functions~\eqref{eqn:velocity_form} as a function of normalized distance $d / d_\text{max}$ for different values of $\alpha$. The vertical axis $V^*(=V/\mathcal{V}_{\text{max}})$ is a normalized velocity by the maximum speed. In general, a smaller value of $\alpha$ results in a larger imposed safety distance.}
    \label{fig:velocityfunctioncurve}
\end{figure}

\begin{figure} 
   \centering
   \subfloat[Linear velocity\label{fig:linear}]{ 
        \includegraphics[width=0.47\linewidth,trim={5cm 2.5cm 2cm 1.5cm},clip]{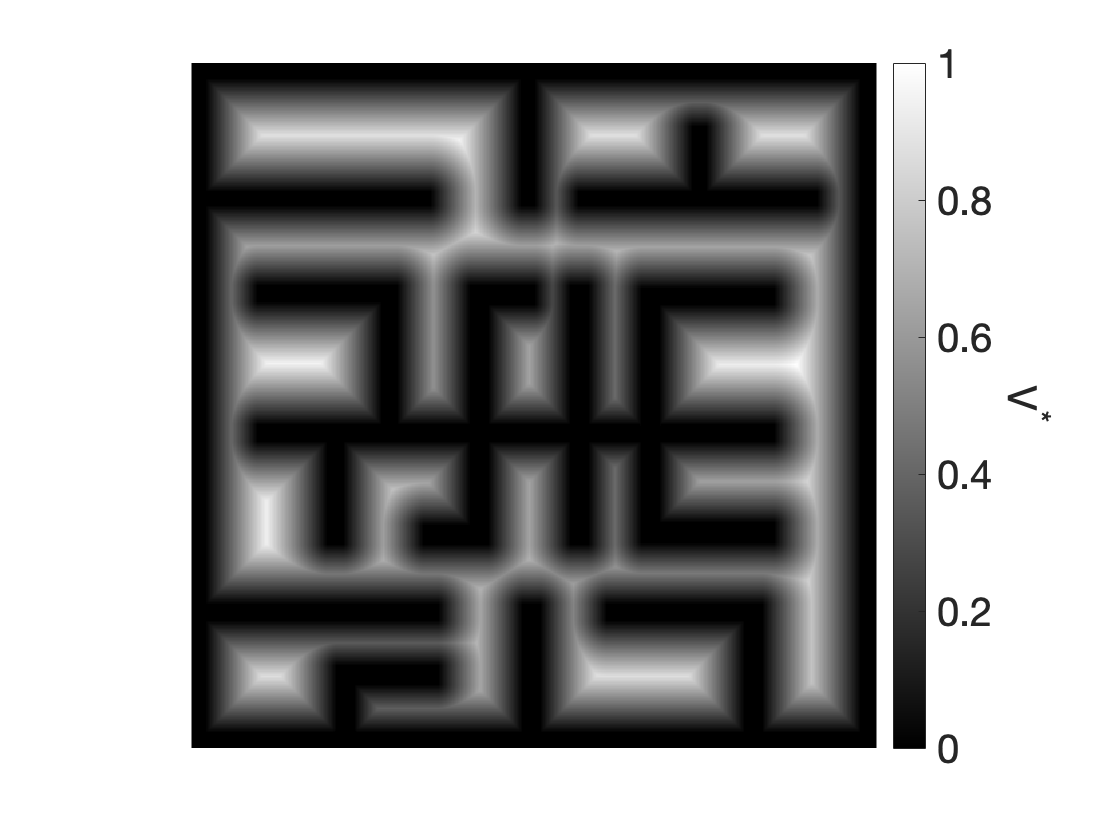}}
   \hfill
   \subfloat[$\alpha=3$ in \eqref{eqn:velocity_form}\label{fig:alpha3}]{%
   \includegraphics[width=0.47\linewidth,trim={5cm 2.5cm 2cm 1.5cm},clip]{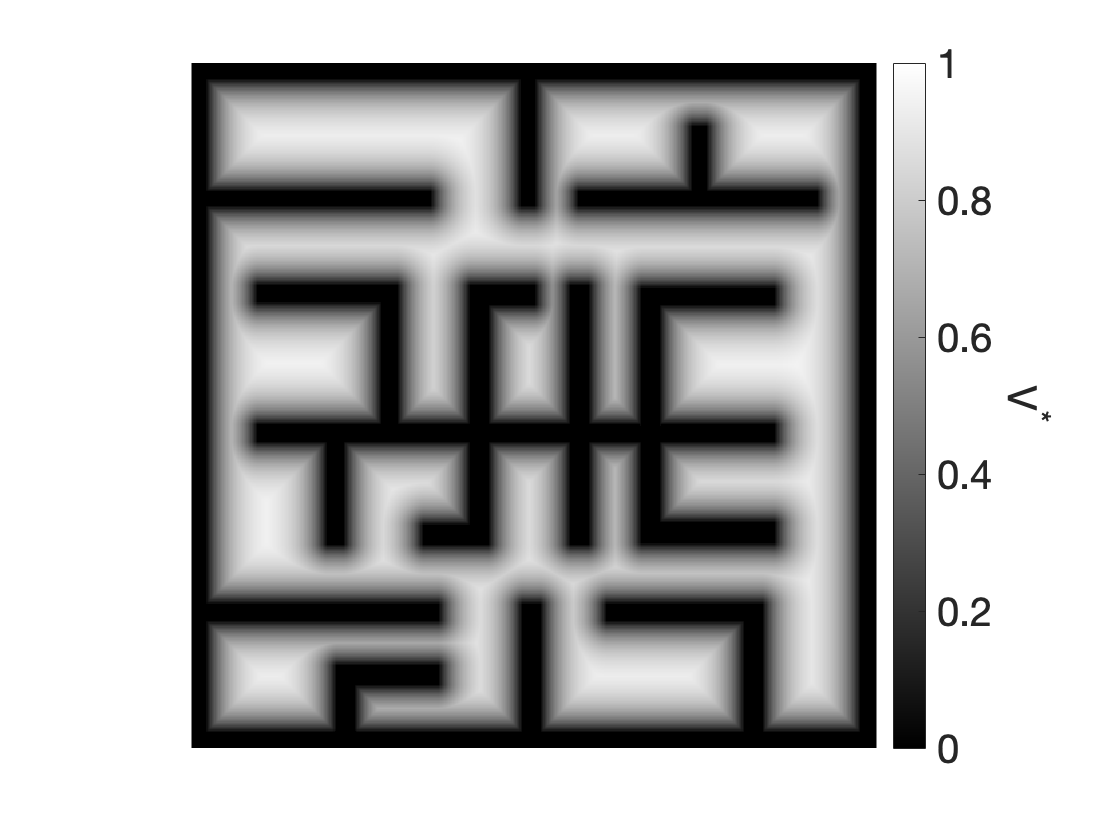}
   }
    \\
    \subfloat[$\alpha=10$ in \eqref{eqn:velocity_form}\label{fig:alpha10}]{%
    \includegraphics[width=0.47\linewidth,trim={5cm 2.5cm 2cm 1.5cm},clip]{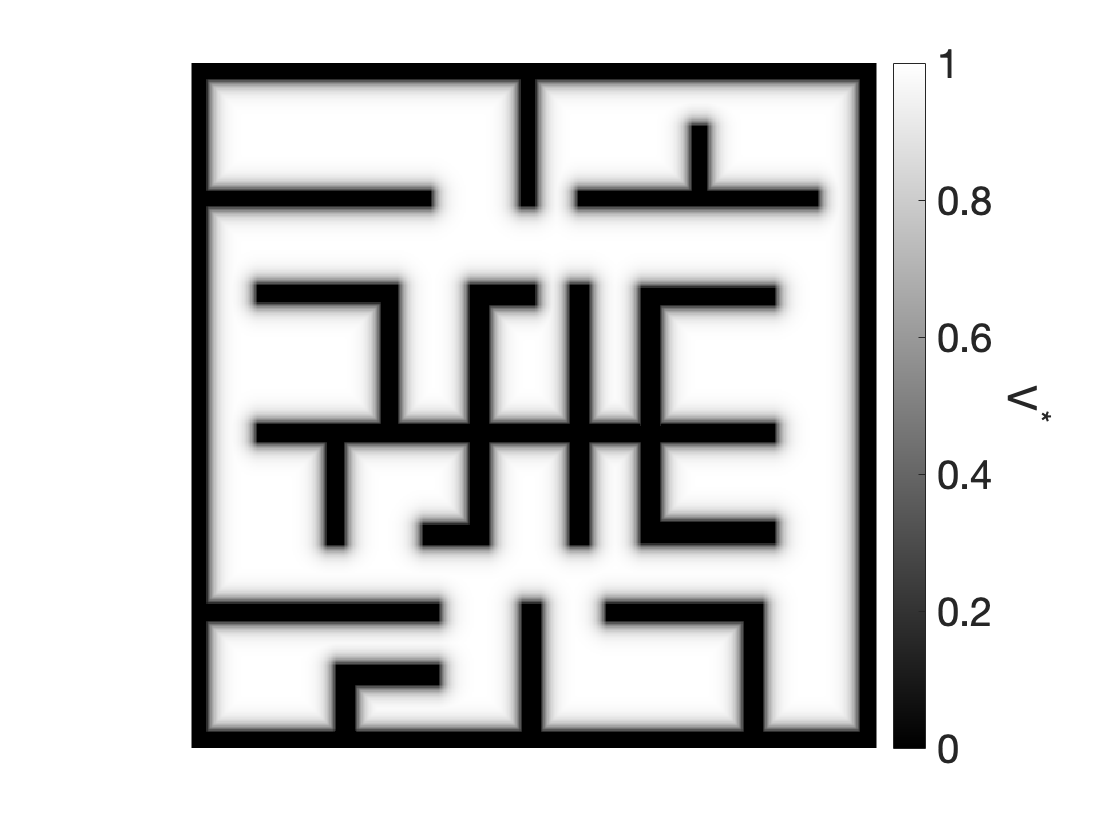}}
    \hfill
    \subfloat[$\alpha=20$ in \eqref{eqn:velocity_form}\label{fig:alpha20}]{%
    \includegraphics[width=0.47\linewidth,trim={5cm 2.5cm 2cm 1.5cm},clip]{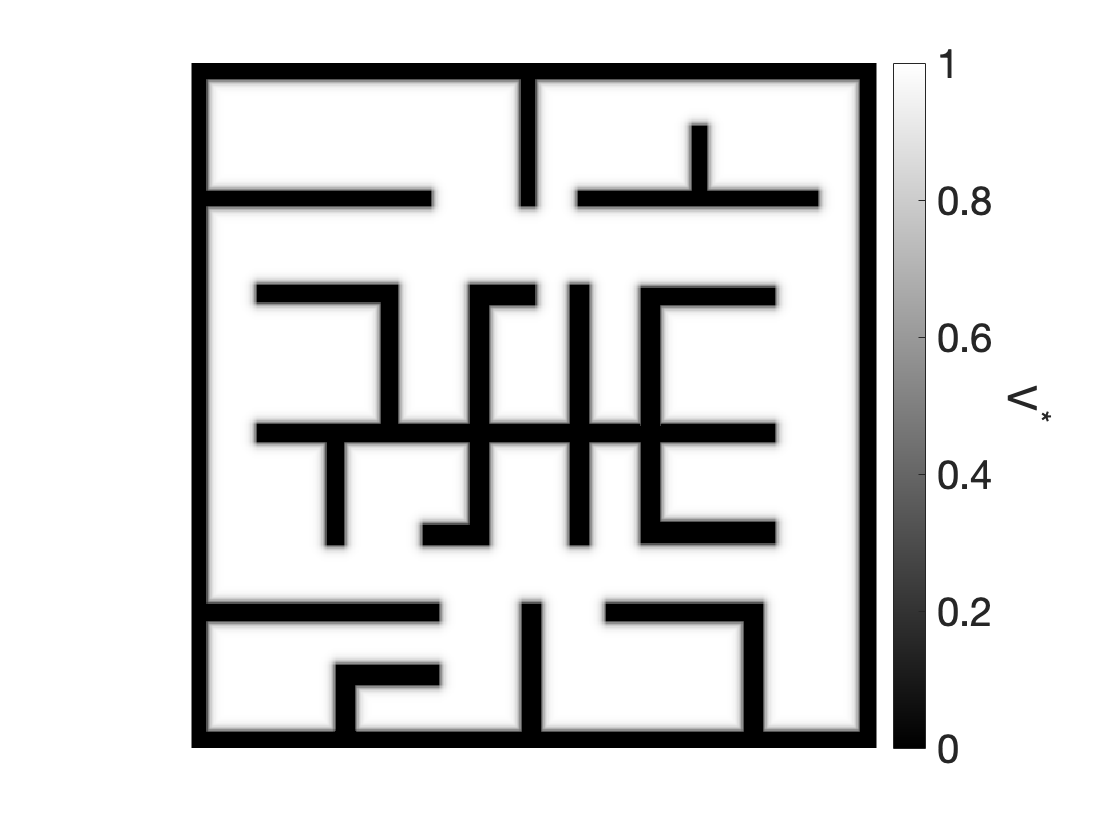}}
   \caption{Comparison of velocity maps generated from the different velocity forms shown in \fref{fig:velocityfunctioncurve}. Sharper increase of $V^{*}$ to value~$1$ results in a larger safety distance.}
   \label{fig:velmap_comparison} 
\end{figure}

\begin{figure}
    \centering
    \includegraphics[width=0.75\linewidth,trim={2.3cm 2.3cm 2.2cm 1.8cm},clip]{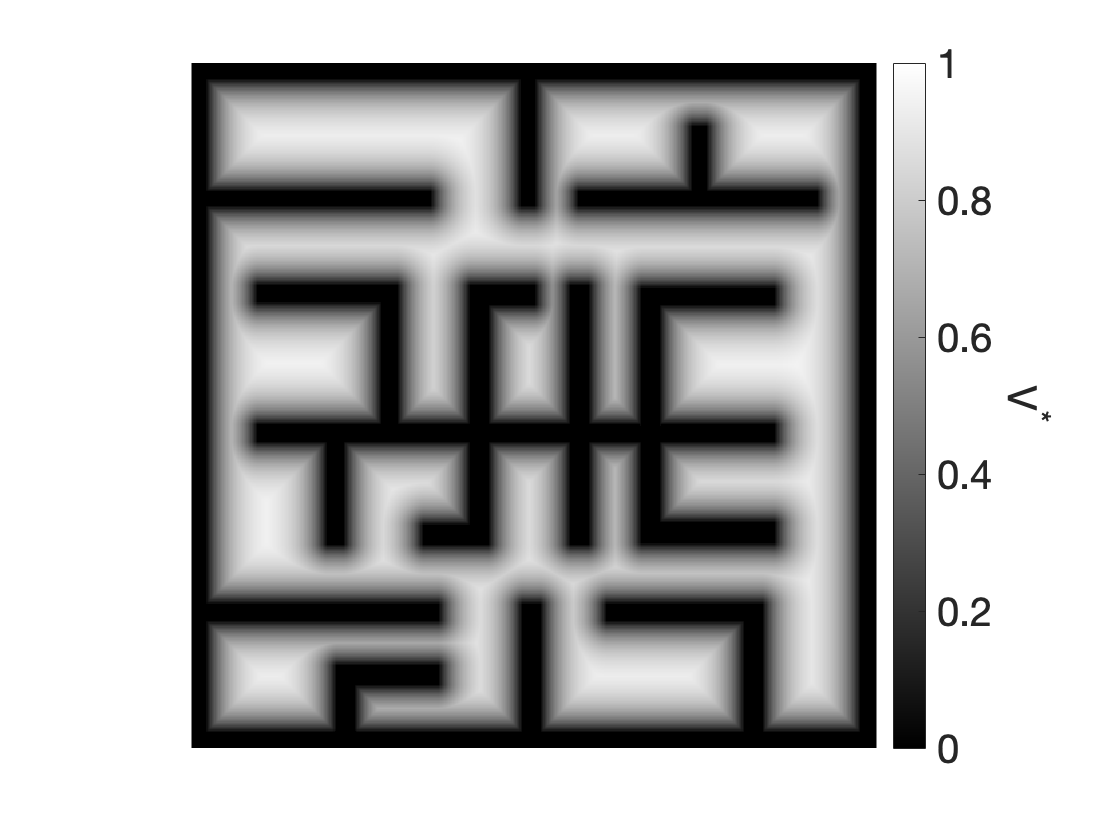}
    \caption{Velocity map created using the form~\eqref{eqn:velocity_form} with $\alpha=3$.
    The velocity values $V^*$ are normalized with the maximum speed of agent $\mathcal V_{\text{max}}$. }%
    \label{fig:normvelmap}
\end{figure}

Alternatively, one may also consider
\begin{equation}
    V(d(\bx)) = \mathcal{V}_{\text{max}} \left[1 - \exp{ \left(-\alpha \left(\frac{d}{d_\text{max}}\right) \right)} \right],
    \label{eqn:velocity_form}
\end{equation}
where $d_\text{max}$ is the maximum distance in the configuration space and $\mathcal{V}_\text{max}$ is the maximum speed of the agent at\change{ }free space, respectively. Note that the form~\eqref{eqn:velocity_form} includes a dimensionless free parameter $\alpha$ that indirectly governs the safety distance. \fref{fig:velocityfunctioncurve} shows the plots of the velocity function profiles at several values of $\alpha$. The corresponding velocity maps for those $\alpha$ values in \fref{fig:velocityfunctioncurve} are shown in \fref{fig:velmap_comparison} to visualize the impact of $\alpha$ values on the collision safety distance. The velocity map created from the binary map (\fref{fig:binaryOccupancymap}),
using the form~\eqref{eqn:velocity_form} with $\alpha=3$, is shown in \fref{fig:normvelmap}. 

\begin{figure}
    \centering
    \includegraphics[width=0.8\linewidth]{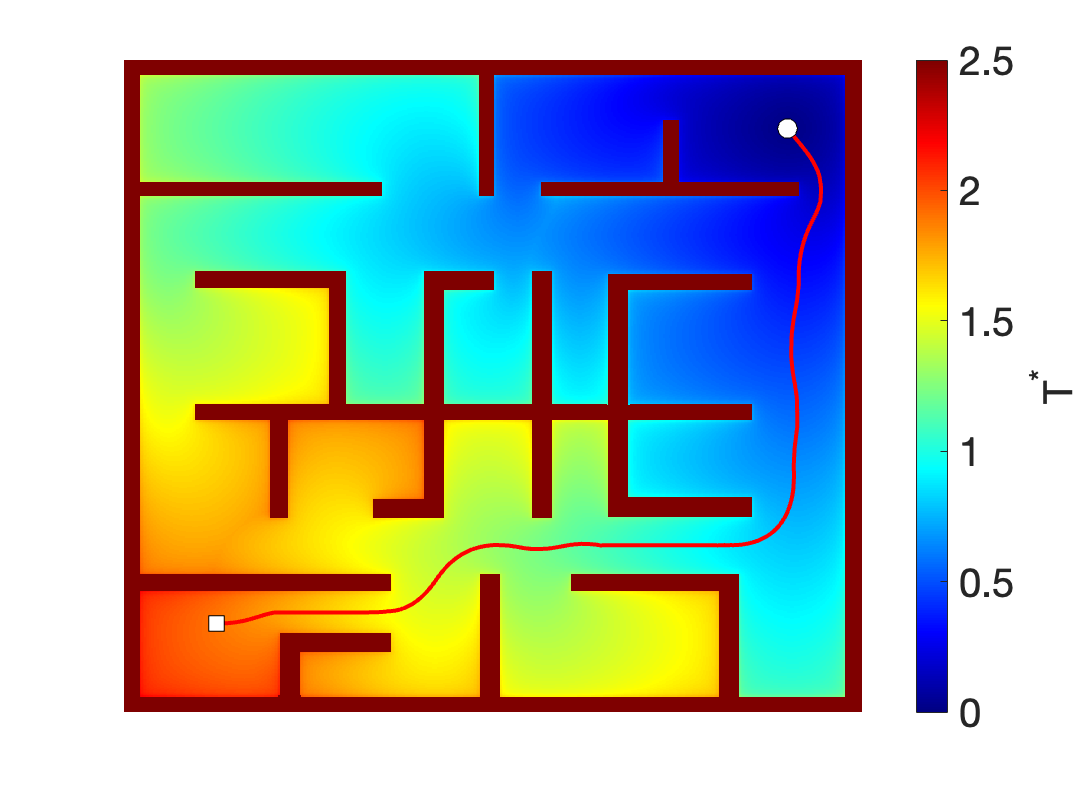}
    \caption{The optimized path after applying the gradient descent algorithm is plotted on the time grid. The white circle denotes the start point, while the square indicates the endpoint.}
    \label{fig:fmsexamplepath}
\end{figure}

Next, the second FMM is executed from the initial position $\bx_0$ of the agent (or vehicle) to compute the time grid map $T(\bx)$, respecting the environmental constraints through $V(\bx)$. 
Finally, the path is obtained again by applying the gradient descent algorithm to $T(\bx)$ and the resulting path for the example case shown in \fref{fig:fmsexamplepath}. The strength of the FMM-based method lies in its unparalleled computational speed when dealing with specific types of optimization problems.
For instance, to provide a more intuitive grasp of the computational efficiency inherent in FMM-based methods, we delve into some practical specifics. The process of extracting a path from a grid of size $10^7$ typically demands only a matter of seconds when employing a single-core machine. To put this into a simpler perspective, it is comparable to handling a two-dimensional pixel image measuring $4000\times4000$ in dimensions. It is also noteworthy that the application of the FMM across multiple iterations does not burden the optimization process with any substantial computational time constraints. The efficiency of the FMM-based method inspires the development of a new framework for various scenarios of modern operations of uncrewed vehicles in the subsequent section.

\section{FMM-based Rendezvous Path Planning for a Team of Heterogeneous Vehicles}
\label{sec:method}

The goal of this section is to introduce an innovative approach to leveraging the FMM-based method within \change{the} multi-agent path planning domain. In particular, we \change{propose} an FMM-based rendezvous path planning algorithm designed for a diverse team of vehicles. The team is tasked with efficiently converging at a single location, aiming for optimal efficiency in pursuit of a general goal.

\subsection{Problem Statement}
This paper considers the problem of finding paths for $N (\geq 2)$ heterogeneous vehicles in a team, which are tasked with rendezvousing within\change{ }minimal time. The region of interest $\Omega$ is assumed to be represented by an occupancy grid map, where each pixel is either free $\mathcal{C}_{\text{free}}$ or occupied $\Omega \backslash \mathcal{C}_{\text{free}}$. According to\change{ }\cite{lavalle2006planning}, a path is viewed as a continuous function $\tau:[0,1]\rightarrow \mathcal{C}_{\text{free}}$, in which each point along the path is given by $\tau(s)$ for some $s \in [0,1]$. Here, $\tau(0)$ corresponds to the starting point of the agent whereas $\tau(1)$ denotes the target point. Although the orientation of each vehicle \change{is} not considered in this work, it is also feasible to incorporate their orientations using the existing methodology~\cite{LiuUSVHeadingAngle16}. 

We assume that the starting position $\tau^{i}(0)$ of each vehicle in the team is given. Note that we introduced the index $i=1,...,N$ to denote each vehicle. Then, the rendezvous path planning for the team is divided into two sub-problem. The first problem is to determine the optimal rendezvous point $\bx_m$ such that 
\begin{equation}
    \bx_m = \operatornamewithlimits{arg\; min}_{\substack{\bx \in \mathcal{C}_{\text{free}}}} \mathcal{F}(\bx), 
\end{equation}
where $\mathcal{F}$ is a general cost function. The second sub-problem is to determine the optimal path \JKadd{$\tau^i$} from \change{the} initial point \JKadd{$\tau^i(0)$} of each agent to the optimal point $\tau^i(1)=\bx_m$. 

From now on, for the purpose of illustration, we fix the optimizing function $\mathcal{F}$ as the meeting time. In rendezvous tasks, this \change{function} corresponds to the arrival time of last agent, which is written as
\begin{equation}
    \mathcal{F} (\bx) = \max \left[ T^1(\bx), T^2(\bx), \dots, T^N(\bx) \right]. 
    \label{eqn:F_example}
\end{equation}
where $T^i (\bx)$ denotes the arrival time for all $\bx\in \mathcal{C}_{\text{free}}$, which will be also referred to as a time grid \change{from now on}. 

\subsection{The Algorithm}
Now, we describe our approach to the aforementioned rendezvous path planning problem. Considering the different initial position\change{s} of each agent, a single implementation of the FMS method\change{ }yield\change{s} $T^i (\bx)$ for all point\change{s} in $\Omega$. During this step, one can consider the arrival time $T^i(x)$ of each agent can be determined considering the different velocities and the safe distances imposed by environments. Once the arrival time maps for all agents are prepared, the \change{time-optimal} meeting point $\bm x_m$, which minimizes the cost \JKadd{$\mathcal{F}(\bx)$} can be conveniently determined by
\begin{equation}
    \bx_m = \operatornamewithlimits{arg\; min}_{\substack{\bx \in \mathcal{C}_{\text{free}}}} \big( \max \left[ (T^1(\bx), T^2(\bx), \dots, T^N(\bx) \right]   \big)
    \change{ }
    \label{eqn:optimal_point}
\end{equation}

\begin{figure}[t]
    \centering
    \subfloat[\label{fig:exampleagent1}]{
        \includegraphics[width=0.47\linewidth]{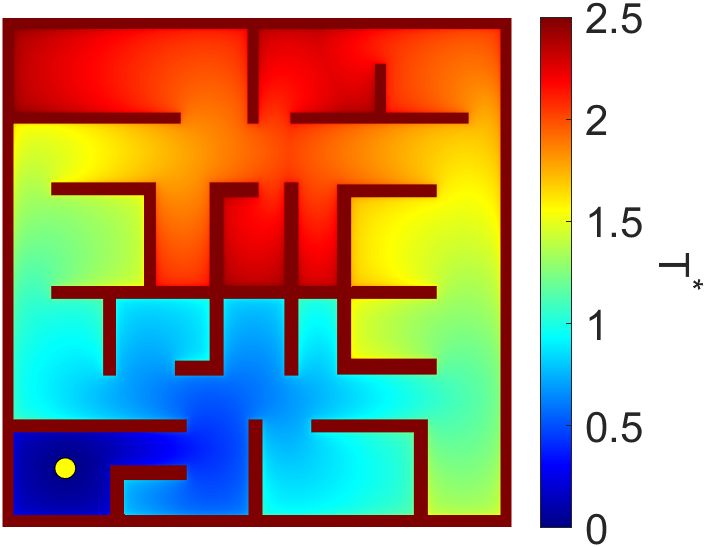}}
    \hfill
    \subfloat[\label{fig:exampleagent2}]{
    \includegraphics[width=0.47\linewidth]{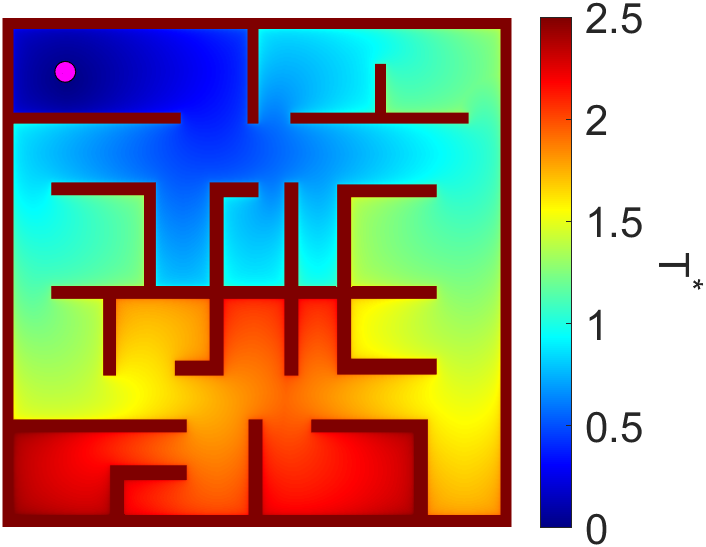}}
    \\
    \subfloat[\label{fig:exampleagent3}]{
    \includegraphics[width=0.47\linewidth]{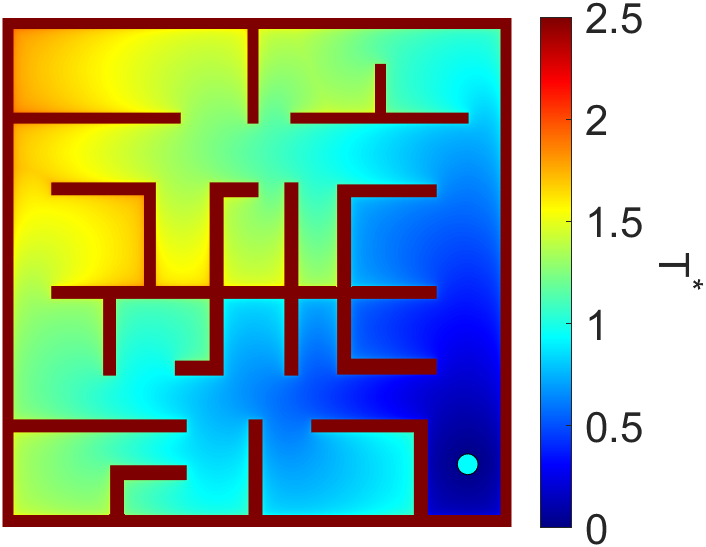}}
    \hfill
    \subfloat[\label{fig:examplepaths}]{
    \includegraphics[width=0.47\linewidth]{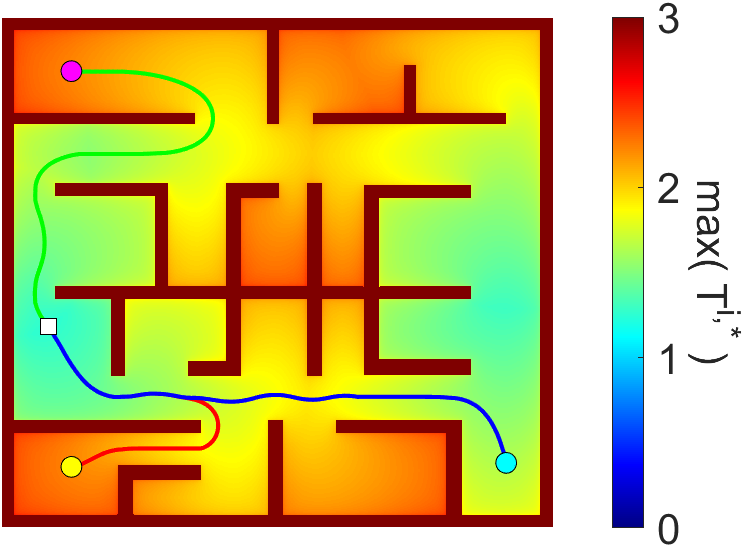}}
    \caption{(a-c) The normalized arrival time $T^*$ maps for three agents located at different initial points. (d) The optimized path drawn on $\mathcal{F}(\bx)$ as defined in the form \eqref{eqn:F_example}.}
    \label{fig:MultiagentFMM} 
\end{figure}

\change{The rest of this section} provides the implementation detail\change{s} of the presented approach using an example of rendezvous planning for three agents, which are initially located at three different corners of a given binary occupancy map previously shown in \fref{fig:binaryOccupancymap}. The initial positions are shown in \fref{fig:exampleagent1}, \fref{fig:exampleagent2}, and \fref{fig:exampleagent3}. For simplicity, we assume that the vehicles are identical, which means that the vehicles travel with the same dynamics and at \change{the} same and constant speed. Specifically, $\alpha =3$ and $\mathcal{V}_{\text{max}}=1$ \change{are} used in the example. 

\HJadd{The algorithm first begins by following the standard step of the FMS method to measure the distance $d\in\mathbb{R}^+$ to the nearest obstacles at every point in the grid}. The first FMM runs from the initial surfaces of obstacles to fill \change{in} $d$-values on every non-occupied point in $\mathcal{C}_{\text{free}} \subset \Omega$, using the uniform velocity $V(\bx) = 1$. Next, we generate a velocity map $V^i(\bx)$ for each agent $i\in\{1,2,...,N\}$ using the velocity function~\eqref{eqn:velocity_form}. Each agent may have a different value of safety parameter $\alpha$ and the maximum allowable speed $\mathcal V_{\text{max}}$. The velocity map for the binary occupancy map using \change{the} form is shown in \fref{fig:normvelmap}. 

Then, we run the second FMM multiple times starting from each initial position of \change{the} agent $\bx^i_0$, which corresponds to $\tau^{i}(0)$. This second round of FMM computation is executed to propagate a source wave point located at the target point until the arrival time $T$ value at the initial point is determined. At each iteration, we obtain the arrival time\change{ }map $T^i(\bx)$ for each agent. Once the iterations of the second FMM \change{are complete}, the optimal point $\bx_m$ can be determined directly from the form~\eqref{eqn:optimal_point}. The result of the term in~\eqref{eqn:optimal_point}, $\max \left[ (T^1(\bx), \dots, T^N(\bx) \right]$, is shown with a color map in \fref{fig:examplepaths}. 

\begin{algorithm}[t]
    \caption{FMM-based algorithm of path optimization for multi-agent rendezvous tasks}
    \label{al:fmm_for_multiagent}
    \begin{algorithmic}[1]
        \renewcommand{\algorithmicrequire}{\textbf{Input:} }
        \renewcommand{\algorithmicensure}{\textbf{Output:} }
        \REQUIRE  A binary occupancy map $\Omega=\mathcal{C}_{\text{free}}\cup \mathcal{C}_{\text{free}}^c$, a cost function $\mathcal{F}(\bx)$, positions of obstacles $\bx_{\text{obs}}$ and initial points of total $N$ agents $\tau^i(0) \in \Omega$
        \ENSURE The best point $\bx_m \in \mathcal{C}_{\text{free}}$ that optimize the cost $\mathcal F$, and the optimized path $\tau^i$ for each agent

        \STATE Execute the fast marching method from $\partial \mathcal{C}_{\text{free}}^c$ to compute the minimum distance $d(\bx)$ from obstacles\change{.}
        \FOR {$i = 1$ to $N$}
        \STATE Calculate the maximum velocity field $V^i(\bx)$ using $d(\bx)$\change{.}
        \STATE Execute the fast marching method from $\bx^i_0$ to compute $T^i(x)$\change{.}
        \ENDFOR

        \STATE Determine the optimizing point $\bm x_m$ that minimizes $\mathcal{F}(\bx)$\change{.}

        \FOR{$i = 1$ to $N$} 
        \STATE \change{U}se the maximum gradient descent algorithm to determine the path $\tau ^i$ from $\bx^i_o$ to \change{\mbox{$\bm x_m$}}.
        \ENDFOR

    \end{algorithmic}
\end{algorithm}

\begin{figure*}[h]
    \centering
    \includegraphics[width=0.8\linewidth]{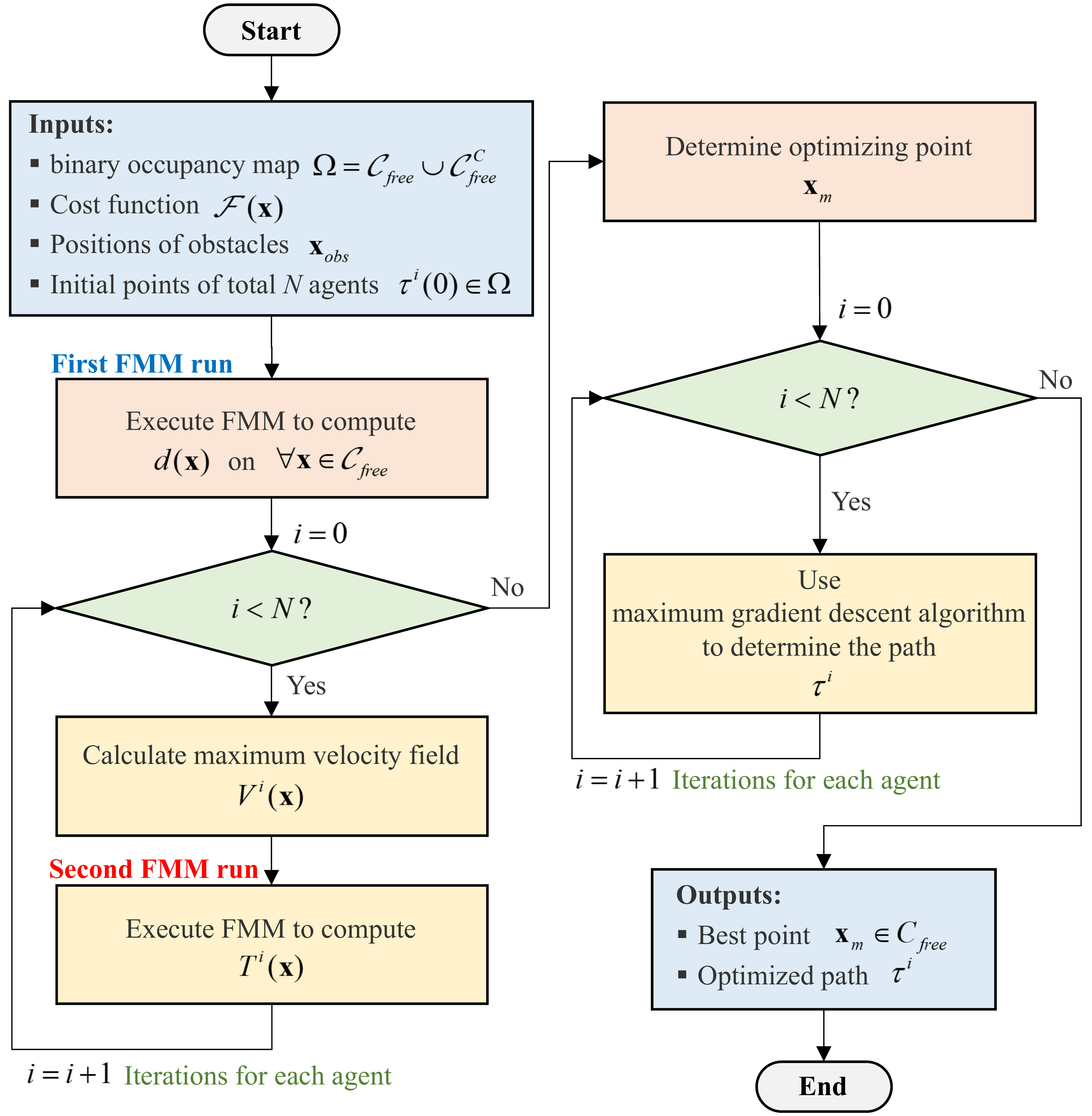}
  \caption{Flow chart for Algorithm~\ref{al:fmm_for_multiagent}, the FMM-based path optimization for multi-agent rendezvous tasks. The algorithm consists of two FMMs.
  The objective of the first FMM is to construct a velocity grid map that takes into account the presence of obstacles, while the second FMM  computes the time grid map $T(\bx)$, respecting the environmental constraints through $V(\bx)$. }
  \label{fig:flowchart}
\end{figure*}

Lastly, the optimized path $\tau^i$ for each agent to the rendezvous point is determined by applying the gradient descent algorithms to the time grid $T^i(\HJadd{ }\bx)$. This trajectory optimization step is inferred from the maximum gradient direction of $T(\bx)$. The final outcome of the FMS method is the optimized continuous path $\tau$, a collection of point in $\Omega$ that guides trajectory of agents as shown in \fref{fig:fmsexamplepath}. The procedure is summarized in Algorithm \ref{al:fmm_for_multiagent}\HJadd{. The corresponding flowchart is shown in \mbox{\fref{fig:flowchart}}, which visualizes the two-step FMM procedure more clearly}.
\JKadd{There are various repositories that can be used to implement Algorithm \mbox{\ref{al:fmm_for_multiagent}} in various programming languages. In our study, the code is developed using a GitHub library source \mbox{\cite{CodeRef}}, which is written in C++ and relies on Boost libraries\mbox{\cite{BOOST}}.} 

\subsection{Algorithm Efficiency and Adaptability}
\label{subsec:efficiency_and_adaptability}

\JKadd{In this section, we test the performance of the suggested algorithm under various conditions. This section includes scalability tests in terms of the number of grids (pixels) and the number of agents in the system. In addition, we demonstrate the additional potential of the present algorithm, which can adapt the path on the fly to reflect environmental changes. All performance tests in this section were carried out on a single 1.6 GHz core with 8 GB of RAM, and we recorded the wall-clock time required to complete one full time step for the two methods.} 

\begin{figure}
    \centering
    \includegraphics[width=0.8\linewidth]{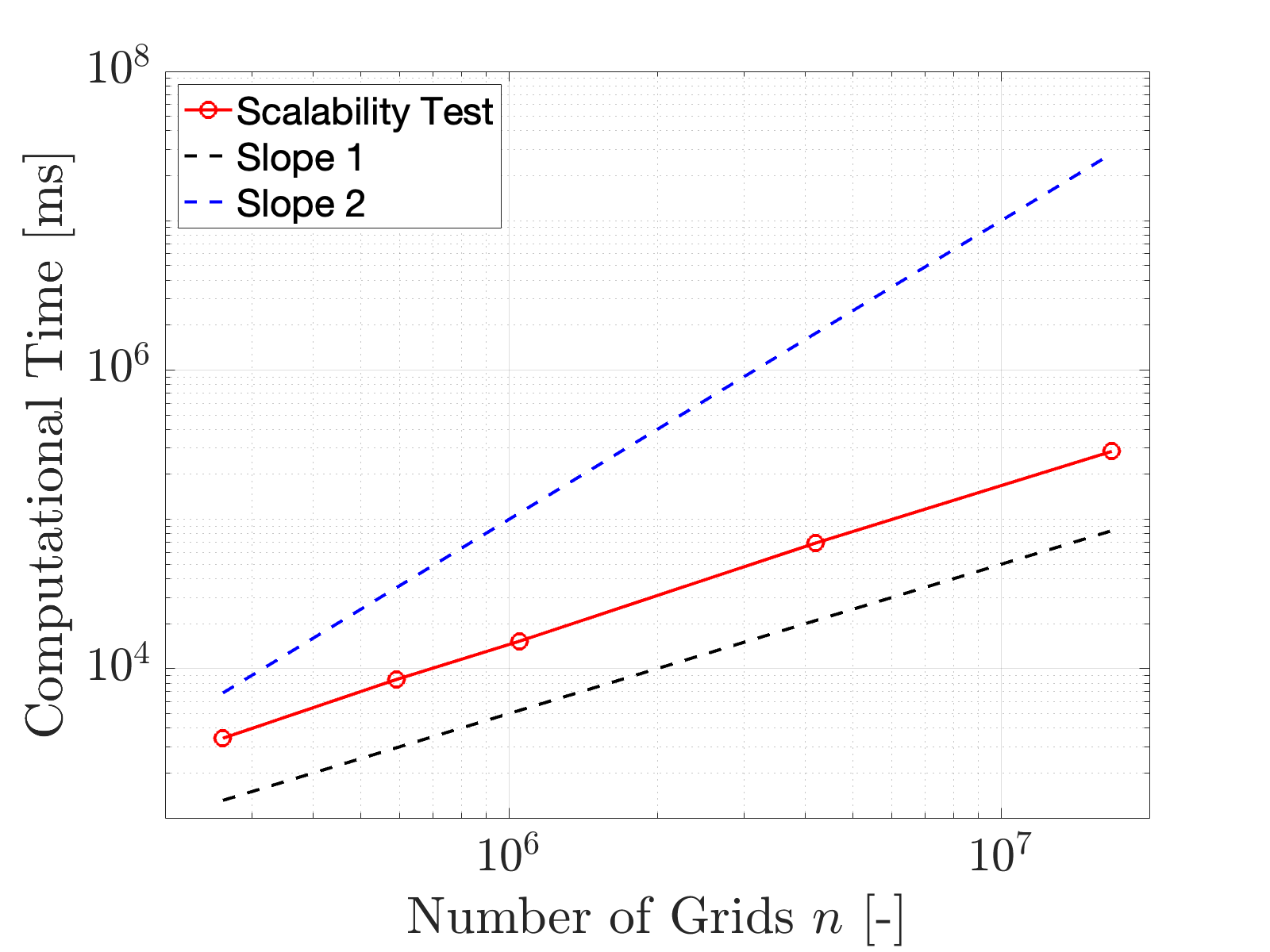}
    \caption{Computational time measured at different grid sizes}
    \label{fig:gridscaletest}
\end{figure}

\JKadd{First, to measure the performance of the algorithm, we conducted the same task (i.e., the rendezvous mission of three agents demonstrated in \mbox{\fref{fig:MultiagentFMM}}), but with different grid sizes. Note that although we did not alter the shape of the map during this test, the increase in the number of grids reflects the size of the environment that can be accessed while maintaining the same resolution. The result is summarized in the format of a log-log scale plot in \mbox{\fref{fig:gridscaletest}}. At the highest grid numbers (\mbox{$4096^2= 16,777,216$}), the actual time it takes is approximately 285 s. While the theoretical (and conservative)  computational complexity is \mbox{$\mathcal{O}(n\log n)$}, in practice we achieve a better scale, approaching \mbox{$\mathcal{O}(n)$}, which appears as a line with slope 1 in the log-log scale plot. The results suggest that the current method can be readily extended to accommodate a larger environment.} 

\begin{figure} 
    \centering
    \subfloat[Rendezvous path planning with total 8 number of agents~\label{subfig:8agents_rendezvous}]{  
    \includegraphics[width=0.8\linewidth]{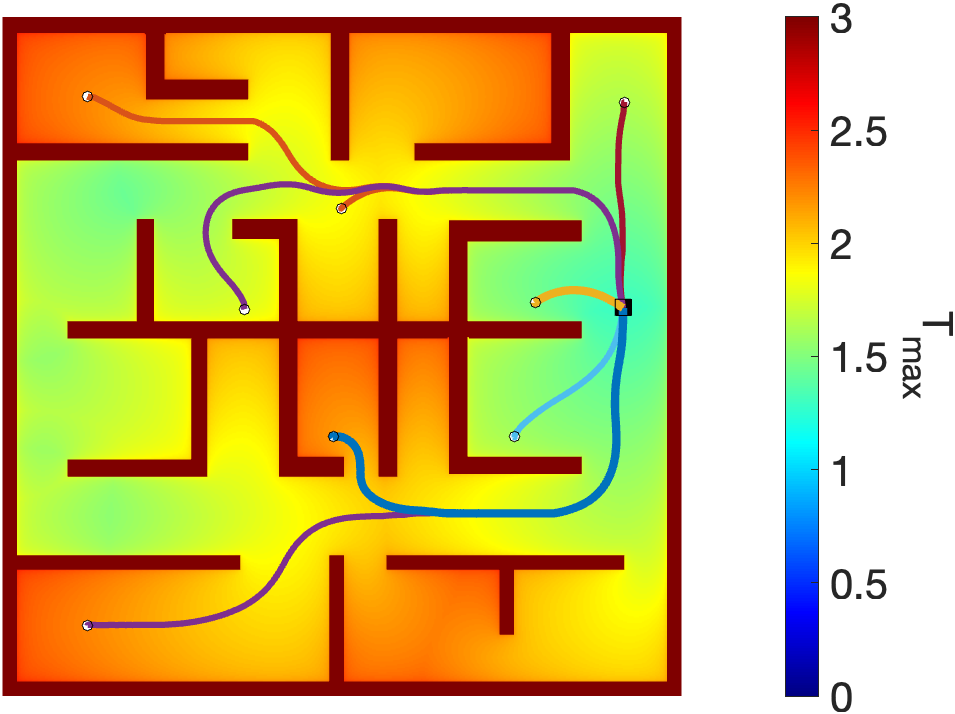}}
    \hfill
     \subfloat[Computational time measured with different numbers of agent~\label{subfig:compute_time_number_agent}]{%
    \includegraphics[width=0.8\linewidth,trim={0cm 0cm 0cm 0cm},clip]{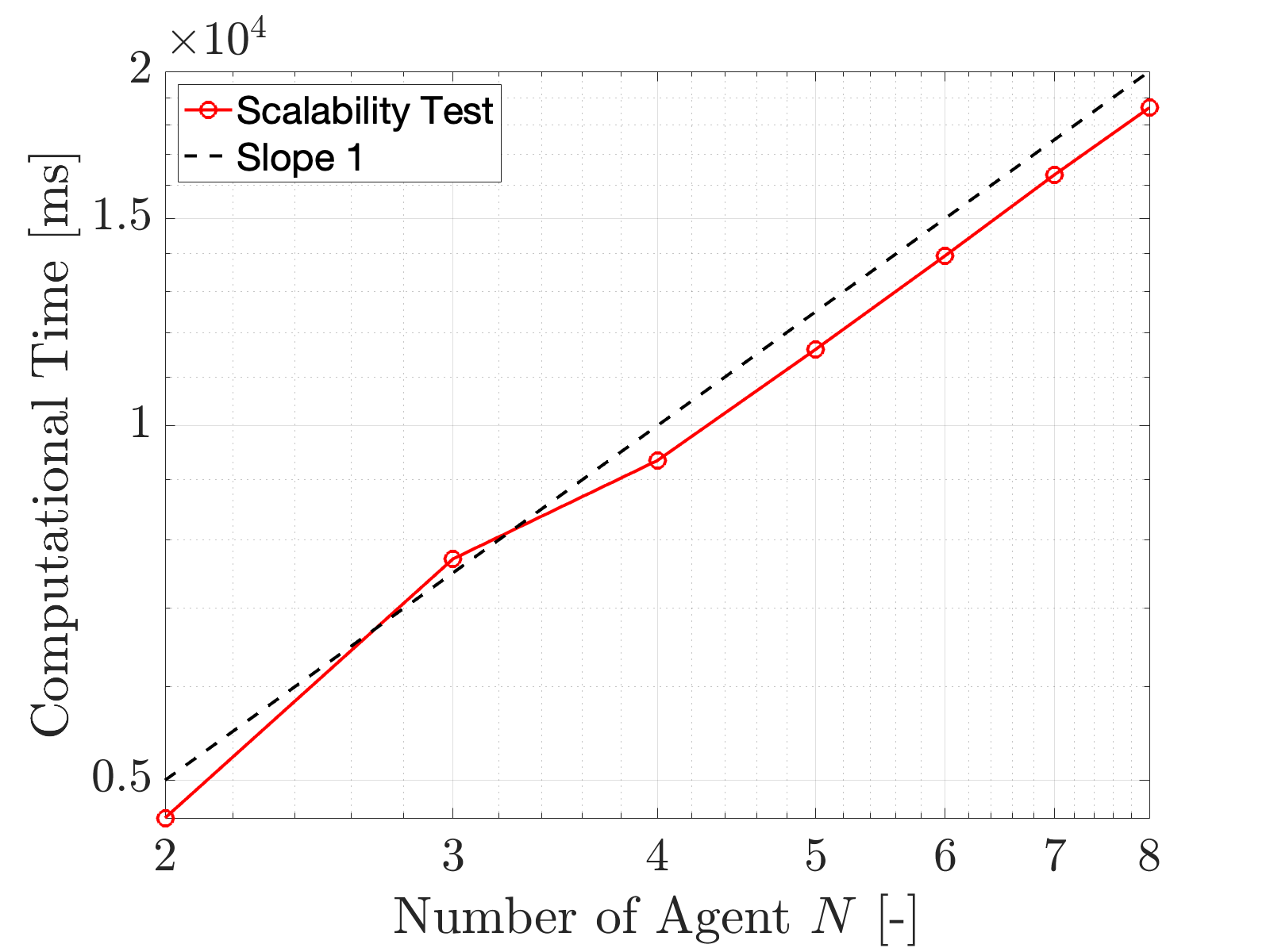}}
    \\
    \caption{Scalability test results on the number of agents}
    \label{fig:scalability_numagents} 
\end{figure}

\JKadd{Next, we also examine how the number of participating agents in the rendezvous task affects the computational time. For example, \mbox{\fref{subfig:8agents_rendezvous}} demonstrates the rendezvous point and path of each agent for a team consisting of 8 agents. While the resulting path planning for all 8 agents seems non-trivial, the increase in the number of agents does not significantly complicate our framework, and the computational time increases only linearly with the number of agents, \mbox{$\mathcal{O}(N)$}, as illustrated in \mbox{\fref{subfig:compute_time_number_agent}}. Moreover, because the computations for each agent are independent, the algorithm can be readily parallelized for further speedup.} 

\begin{figure*}[htbp]
    \centering
    \setkeys{Gin}{width=0.45\textwidth}
    \subfloat[
        \label{subfig:onlinePathPlan_a}]{
        \includegraphics{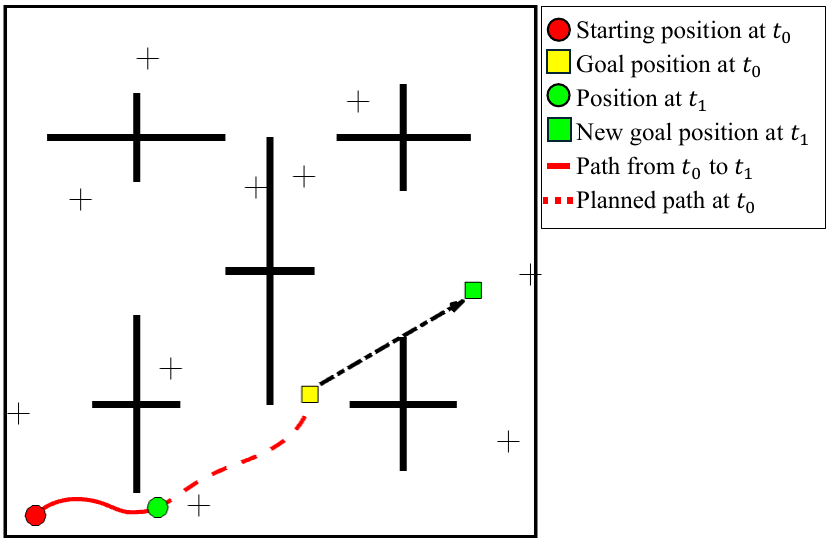}}
          \hspace{1cm}
    \subfloat[
        \label{subfig:onlinePathPlan_b}]{
        \includegraphics{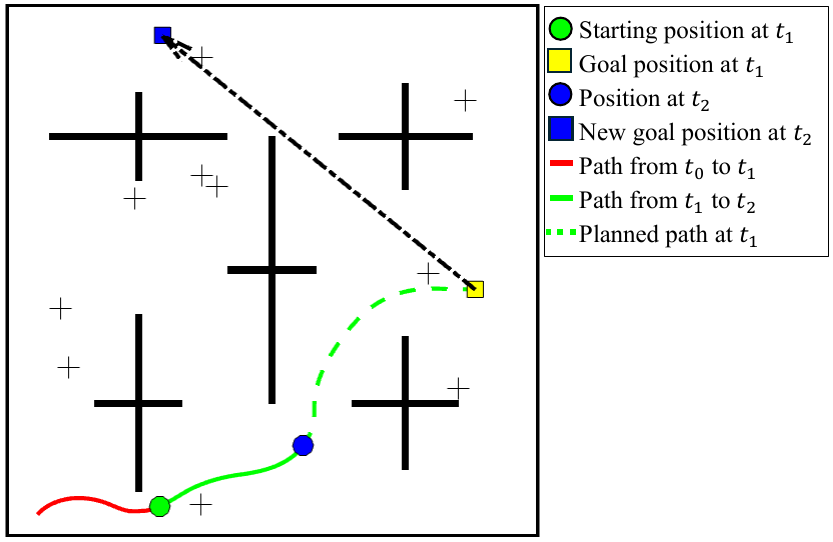}}
    \\
    \subfloat[
        \label{subfig:onlinePathPlan_c}]{
        \includegraphics{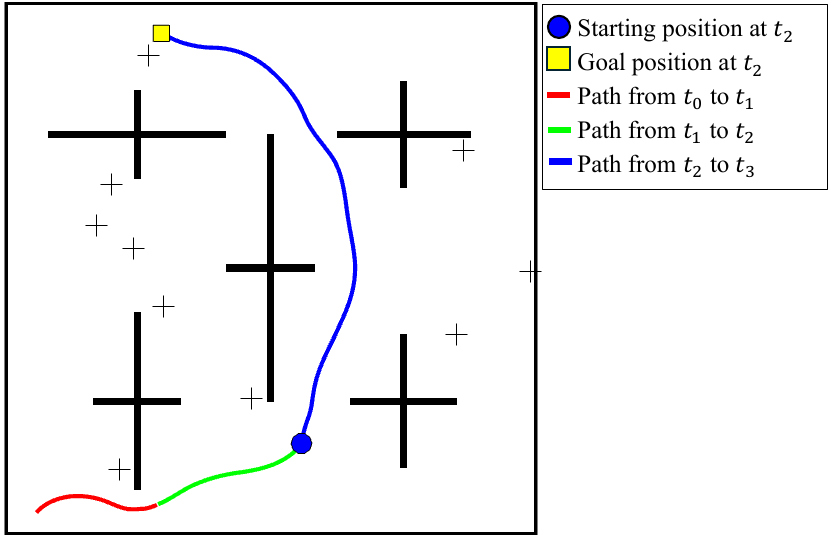}}
          \hspace{1cm}
    \subfloat[
        \label{subfig:onlinePathPlan_d}]{
        \includegraphics{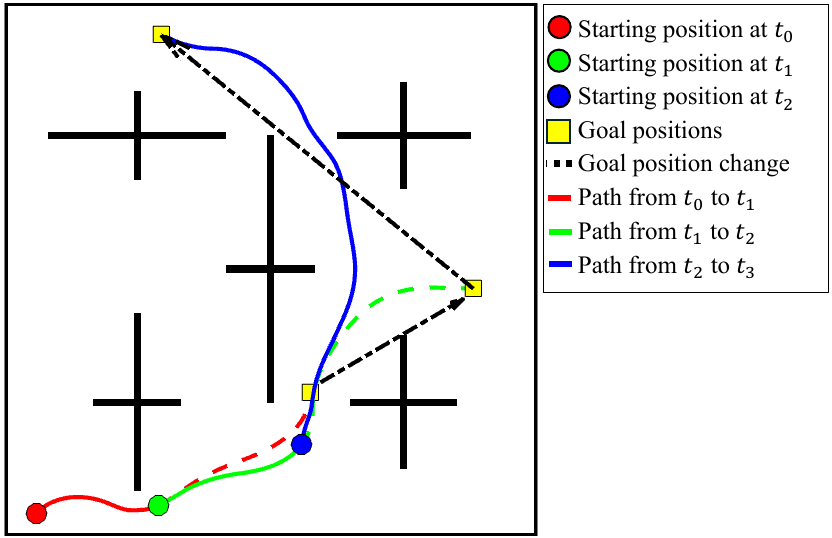}}      
    \caption{An example of semi-discretized online path planning using the FMM method.
    (a) The initial path planning at the first time interval $[t_{0}, t_{1}]$. The red circle and yellow square denote the initial positions of the agent and target point respectively. When the positions of the target and obstacles (black cross points) are updated after $\delta t=200$ [s], a new path is generated. (b) and (c) demonstrate path planning for the second and third time interval $[t_{1}, t_{2}]$ and $[t_{2}, t_{3}]$. (d) Summary the sequence of path planning occurring over three time intervals. In this example, $t_{0}=0$, $t_{1}=\delta t$, $t_{2}=2\delta t$, and $t_{3}=\infty$ are used.}
    \label{fig:onlinePathPlan}
\end{figure*}

\JKadd{Before concluding this section, we briefly show an additional adaptability of the proposed framework for online path planning, leveraging the marginal computation cost (approximately 3 seconds on a 512 × 512-sized regular grid). Specifically, we consider a case where the positions of obstacles and target points are changing over time. Here, we will limit our discussion to cases where the time scale of agents traveling in the domain is much larger than that of the target and environmental changes. In the following, we assume that the positions of agents and moving obstacles are tracked by a system at every time interval $\delta t$.} 

\JKadd{For demonstration, consider an initial path planning using the FMM approach illustrated in \mbox{\fref{subfig:onlinePathPlan_a}}. The initial positions of the agent and target are denoted by the red circle and yellow square, while the black cross markers represent the positions of moving obstacles. After a time interval of \mbox{$\delta t$}, we assume that when the agent arrives at the green circle, the target has moved to the green square. If such environmental changes are recorded, a new FMM-based path planning algorithm, resulting in \mbox{\fref{subfig:onlinePathPlan_b}}, is implemented to reflect the new positions of the target and obstacles. \mbox{\fref{subfig:onlinePathPlan_c}} illustrates an additional path planning for the third time interval, spanning from $2\delta t$ until the goal position is reached. \mbox{\fref{subfig:onlinePathPlan_d}} summarizes the sequence of path planning occurring over three time intervals. The results show that online path planning, which respects environmental variations, can be addressed in a semi-discretized manner due to the extreme computational efficiency of the FMM-based method.} 

\change{Finally, we note that the present approach did not explicitly consider coordination and communication, assuming that the problem's scale was large enough to focus solely on finding the rendezvous point and paths. Specifically, agents' coordination can be considered in detail by extending the implementation to treat other agents as obstacles during the planning stage, as illustrated in \mbox{\fref{fig:onlinePathPlan}}. Communication was assumed to be available throughout the operation, meaning synchronous rendezvous is considered. The method can be implemented in either a centralized or decentralized manner. 
On the other hand, we admit that the approach would fail to find the time-optimal rendezvous path if each agent does not recognize the location of other moving agents. 
Such a situation may occur with only asynchronous communication or, worse, when communication is completely unavailable. If this is the case, we expect that approximated locations (e.g., via dead reckoning) might be utilized to determine the rendezvous point and paths. } 

\subsection{Baseline Comparison}

\begin{figure*}
    \centering
    \subfloat[Rendezvous path planning result from RRT*~\label{subfig:rrt_star}]{        
    \includegraphics[width=0.4\linewidth]{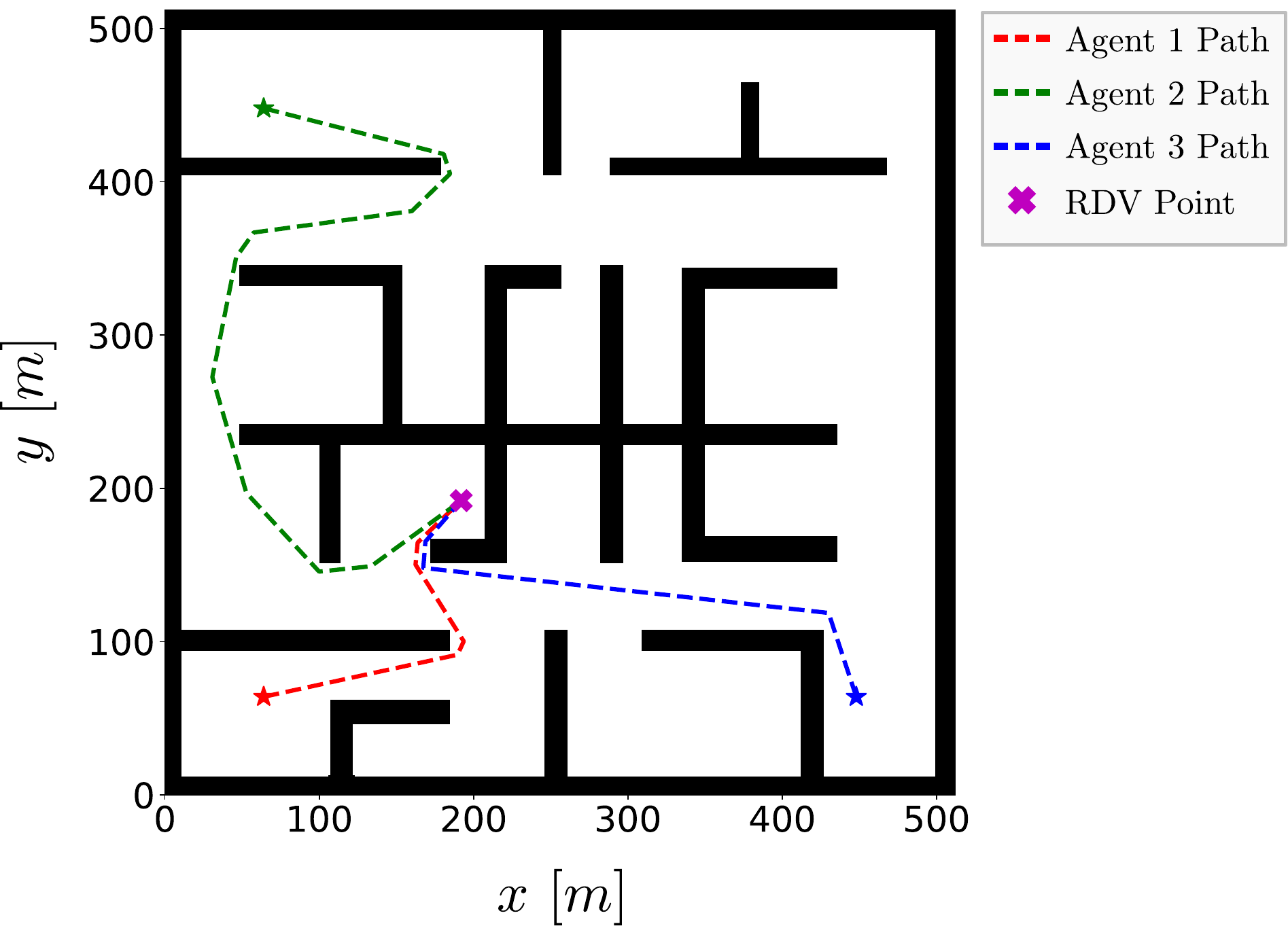}}
    \hspace*{1cm}
     \subfloat[Rendezvous path planning result from the presented FMM-based approach~\label{subfig:fmm_comp}]{%
    \includegraphics[width=0.4\linewidth]{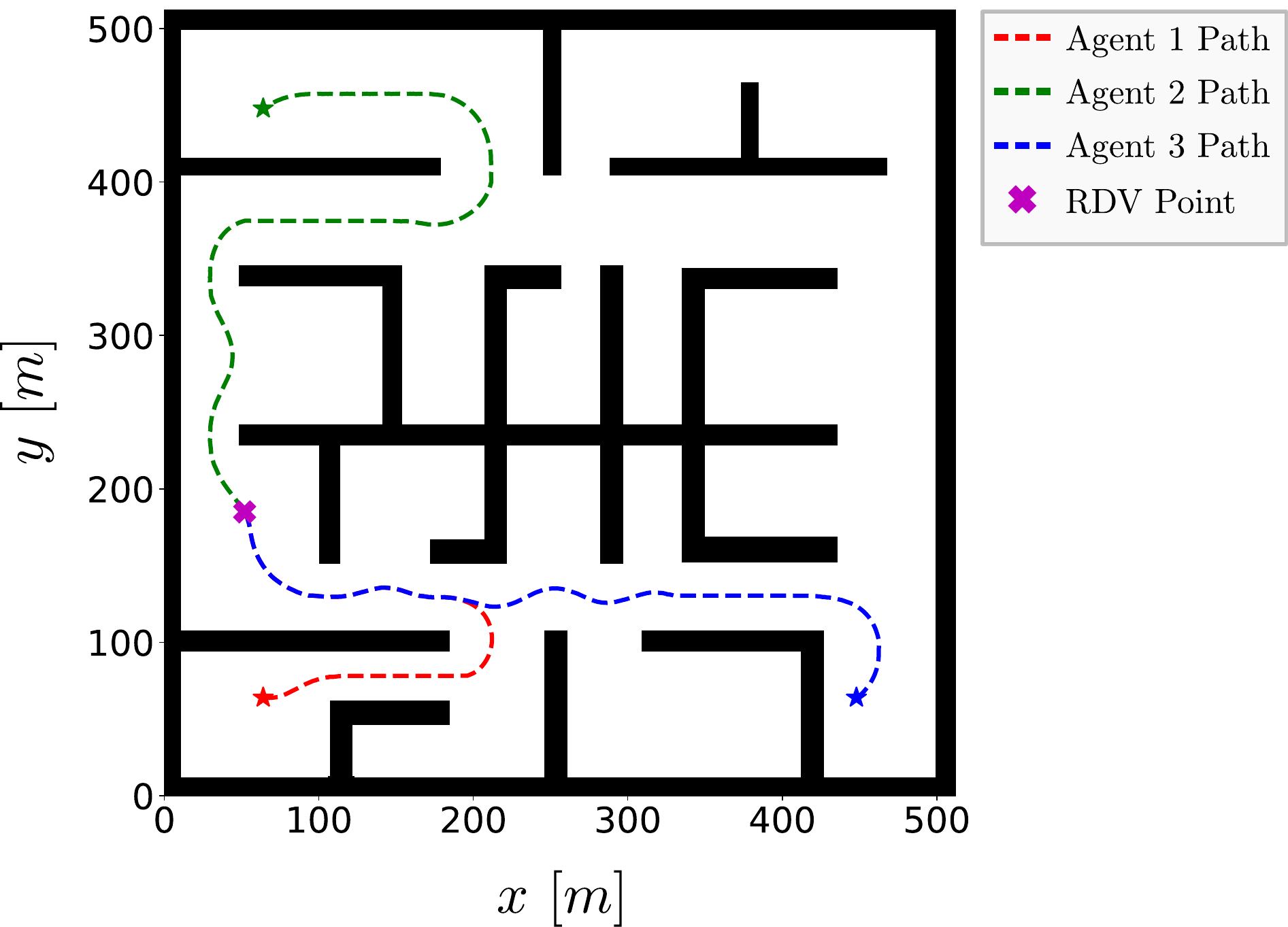}}
    \\
    \caption{Comparison of the presented FMM-based rendezvous path planning method to a baseline method consisting of a weighted centroid method to find the rendezvous point (RDV Point) and the RRT* algorithm (for pathfinding).}
    \label{fig:comparison} 
\end{figure*}

\change{The current approach is compared with a baseline algorithm for evaluation. To the best of our knowledge, as described in detail in Section \mbox{\ref{sec:introduction}}, existing algorithms cannot be directly applied to the problem considered in this paper.
Specifically, scheduling algorithms primarily focus on determining the sequence of heterogeneous agents, rendezvous search algorithms concentrate on identifying the rendezvous position under limited communication, and multi-agent path finding algorithms are chiefly concerned with collision avoidance among agents.
Therefore, we construct a baseline algorithm by combining two methods from different objectives: the weighted centroid point\mbox{~\cite{zeng2014path}} for determining the rendezvous point and the optimal rapidly-exploring random tree (RRT*) \mbox{~\cite{karaman2011sampling}} for path planning. The inverse of each agent's velocity is used as a weight for the centroid calculation, while the RRT* is implemented using the Open Motion Planning Library (OMPL) \mbox{~\cite{sucan2012the-open-motion-planning-library}} for C++ implementation.
The computational complexity of the baseline method is also comparable to the present scheme: the RRT* algorithm has a computational complexity of $\mathcal{O}(\mathrm{K}\log \mathrm{K})$, where $\mathrm{K}$ is the number of nodes in the tree.}

\change{\mbox{\fref{fig:comparison}} illustrates the comparative analysis between the baseline method and the current FMM-based path planning approach using the same example problem considered in \mbox{\fref{fig:MultiagentFMM}}. 
The rendezvous point and agents' paths computed by the baseline method are shown in Fig. \mbox{\ref{fig:comparison}} (a), where it takes 574.29 seconds for all agents to reach the rendezvous. The corresponding results from the FMM method are shown in \mbox{\fref{fig:comparison}} (b), where it takes 626.27 second for all the agents to reach the rendezvous point.}

\change{
At first glance, the advantage of our approach may not seem clear compared to the baseline method. 
However, several points must be considered. First, the RRT* algorithm does not employ any path smoothing operations, so the constructed path may not be convenient depending on the vehicle's kinematics, whereas the FMM path is smooth. Second, the weighted centroid method (for determining the rendezvous point in the baseline approach) does not guarantee that the rendezvous point $\bm{x}_m^{cp}$ remains within $\mathcal{C}_{\text{free}}$. Lastly, if domain $\Omega$ is more complex, resembling a maze, the optimality of the rendezvous point suggested by the weighted centroid method becomes more questionable.
}

\section{Numerical Experiment}
\label{sec:experiment}

In this section, we conduct a numerical experiment \change{to showcase} an application of the suggested method in more realistic cases. We consider a virtual scenario of \change{a} rendezvous task for a team of heterogeneous vehicles. The experimental setting is as follows. 

\subsection{Experimental Setup}

\begin{figure}
    \centering
    \setkeys{Gin}{width=0.32\textwidth}
    \subfloat[
        \label{fig:tampa_sat}]{
        \includegraphics{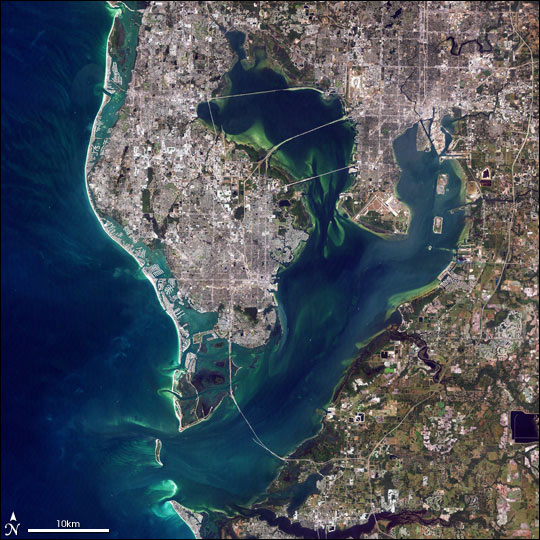}}
    \hfill
    \subfloat[
        \label{fig:binary}]{
        \includegraphics{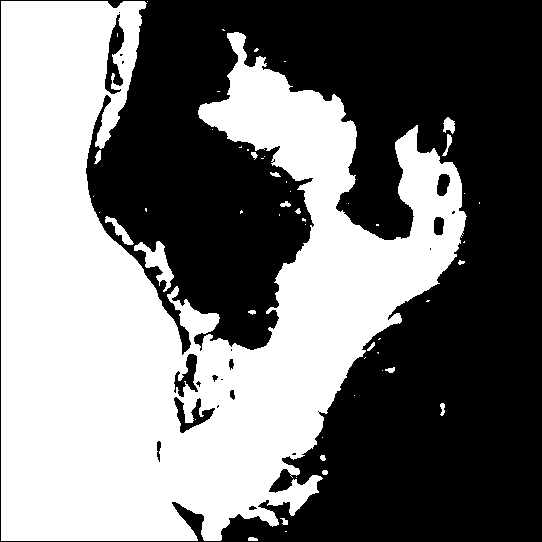}}
    \hfill
    \subfloat[
        \label{fig:reverse}]{
        \includegraphics{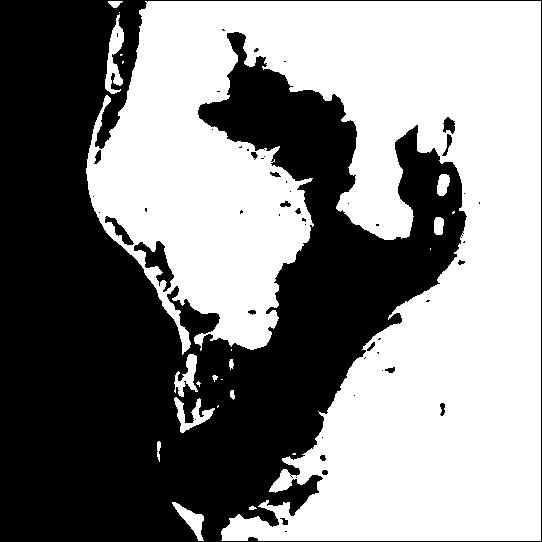}}
    \caption{(a) Satellite image of Tampa Bay, FL. (downloaded from NASA Earth Observatory) (b) Processed binary image from \fref{fig:tampa_sat} for the USV and UUV; (c) Processed binary image for UGV, which is an inverse of \fref{fig:binary}}
    \label{fig:tampa_imageprocessing}
\end{figure}

\begin{figure*}
    \centering
    \subfloat[
          \label{fig:uuv_time}]{\includegraphics[height=0.165\textheight,trim={7cm 2.5cm 8.5cm 1.5cm},clip]{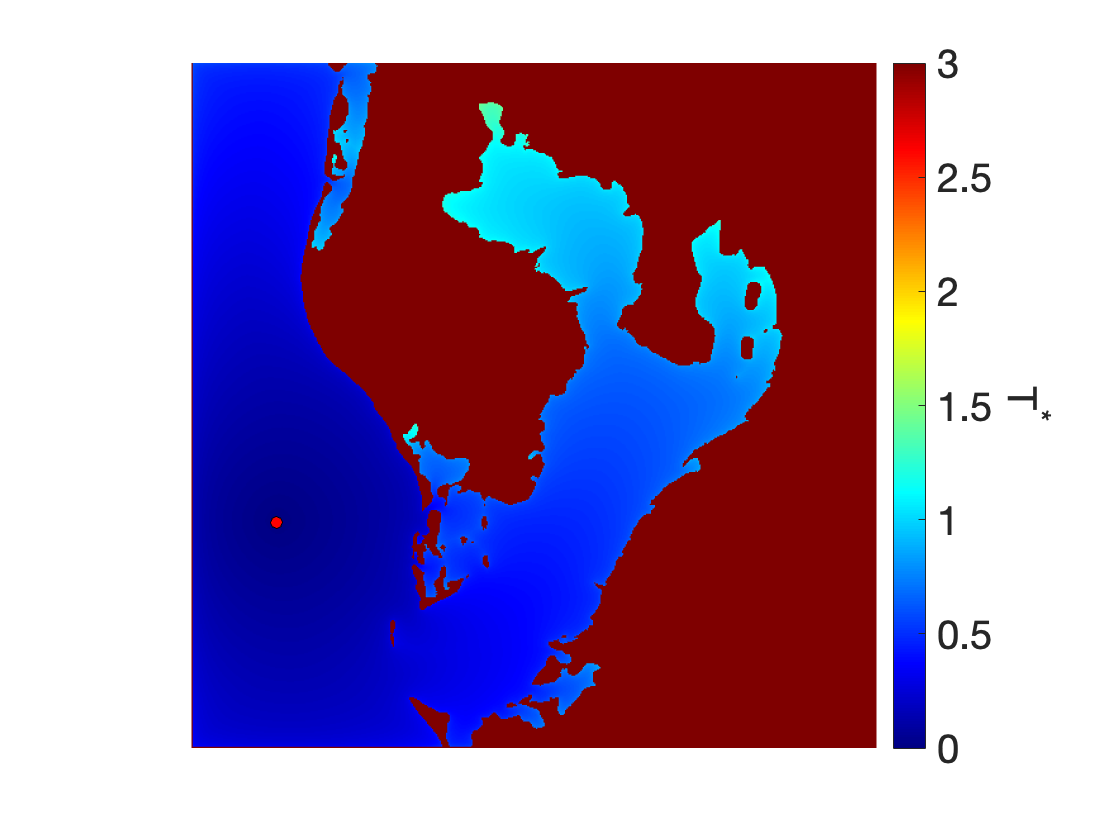}}
    \hfill
    \subfloat[
          \label{fig:usv_time}]{\includegraphics[height=0.165\textheight,trim={7cm 2.5cm 8.5cm 1.5cm},clip]{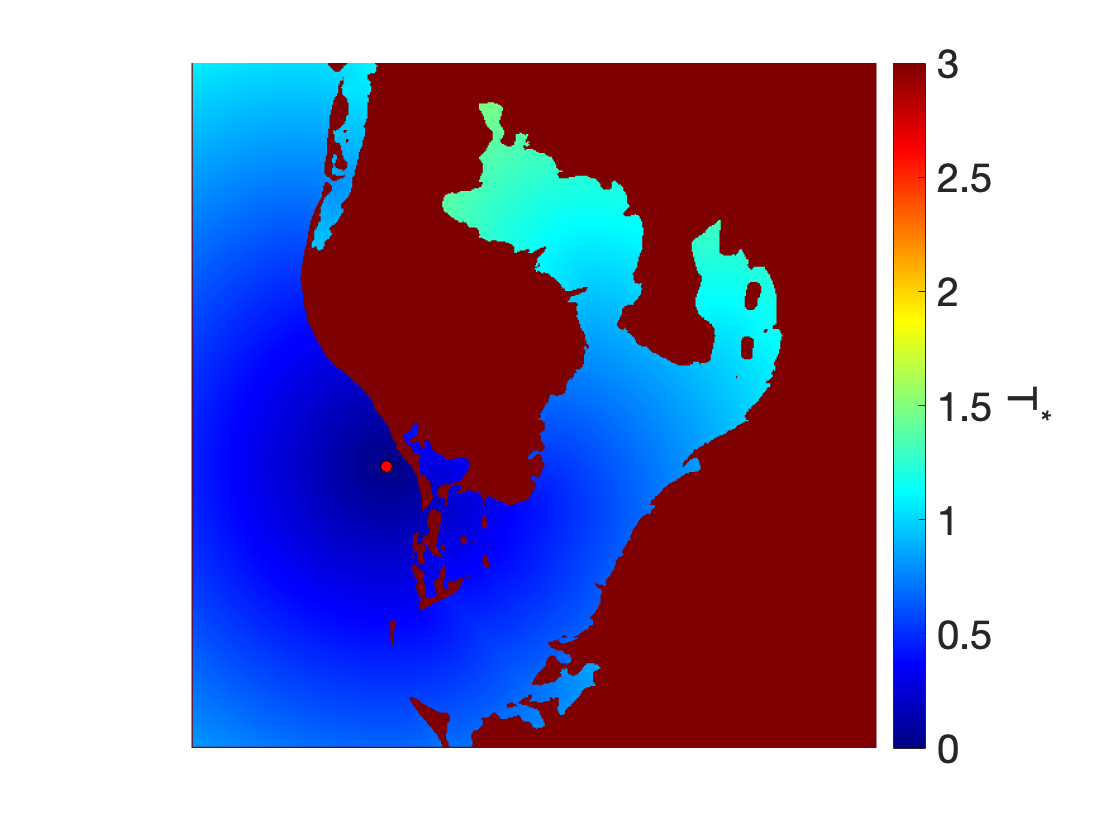}}
    \hfill
    \subfloat[
          \label{fig:ugv_time}]{\includegraphics[height=0.165\textheight,trim={7cm 2.5cm 8.5cm 1.5cm},clip]{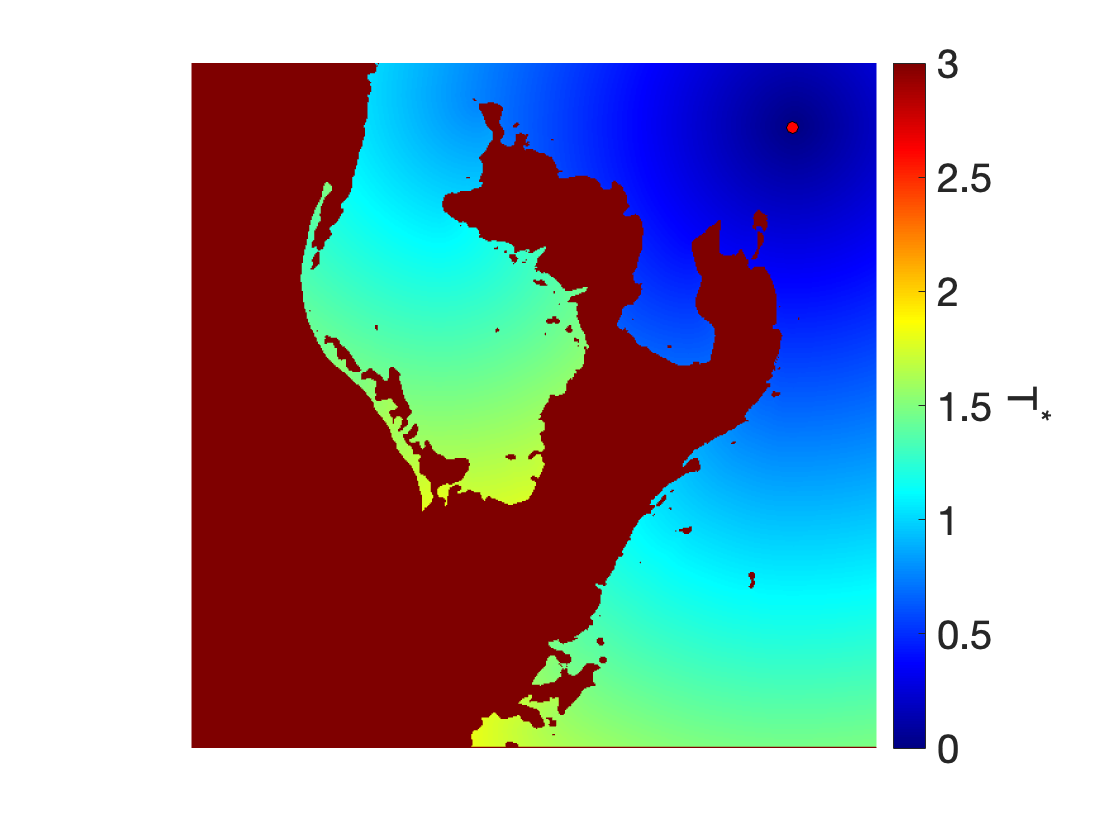}}
    \hfill
    \subfloat[
          \label{fig:uav_time}]{\includegraphics[height=0.165\textheight,trim={7cm 2.5cm 2cm 1.5cm},clip]{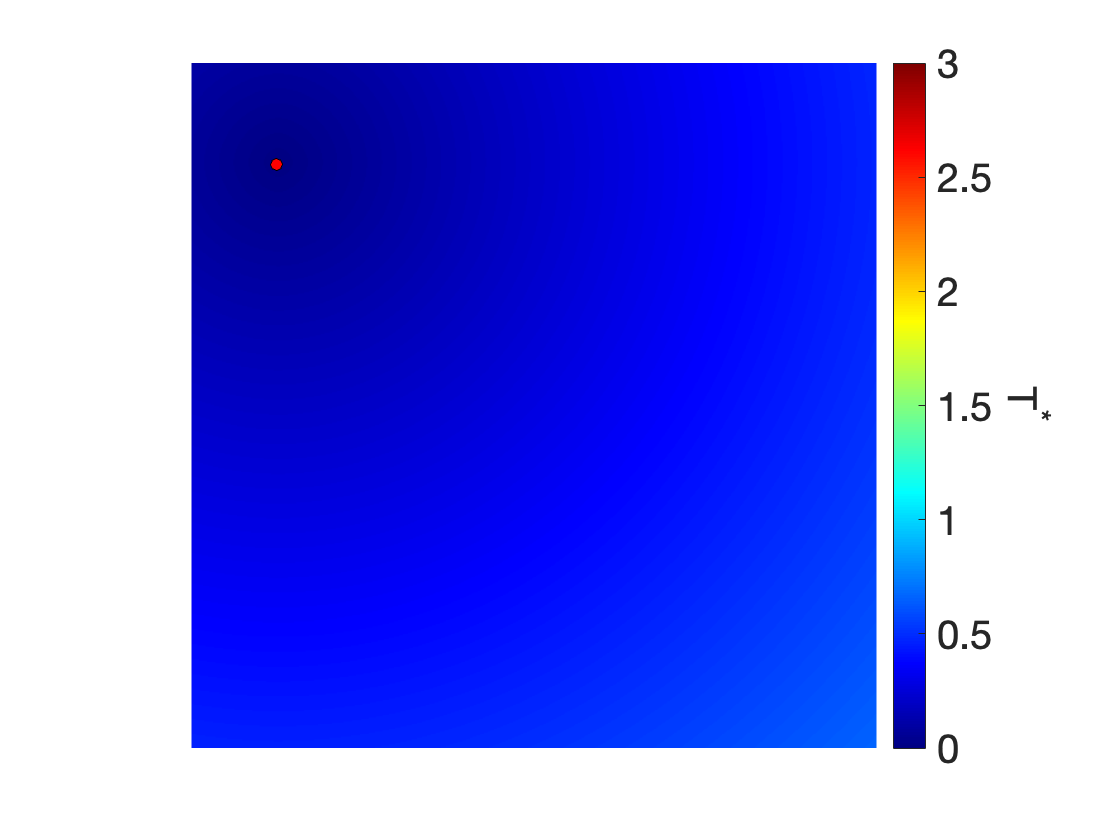}}
    \\
    \caption{The arrival time grids $T^i(\bx)$ of the (a) UUV, (b) USV, (c) UGV, and (d) UAV. The red dot in each plot denote the initial position of the vehicle. }
    \label{fig:time_grids}
\end{figure*} 

First, we create \change{a} computational domain to simulate a realistic environment. The Tampa Bay area \change{is chosen} as our test domain. We use\change{ }a satellite image from NASA's Earth Observatory (as shown in \fref{fig:tampa_sat})\footnote{\url{https://earthobservatory.nasa.gov/images/4745/tampa-bay-florida}}. \change{Then}, the GRIP tool (\textit{Graphically Represented Image Processing engine})~\cite{LeitschuhClark2016} \change{is} employed to convert the satellite image into a binary configuration space map. The primary objective of image processing at this stage \change{is} to distinguish water bodies and land areas, as illustrated by white and black pixels\change{,} respectively in \fref{fig:binary}. 

Next, we buil\change{d} a team of heterogeneous agents, \change{consisting} of\change{ }four types \change{of vehicles}: an uncrewed underwater vehicle (UUV), an uncrewed surface vehicle (USV), an uncrewed ground vehicle (UGV), and an uncrewed aerial vehicle (UAV). The UUV operates exclusively underwater but is limited by operational depth constraints. Consequently, UUV operations are required to take place at a considerable distance from the shoreline. On the other hand, the USV is designed for slower mobility, but it has the capability to navigate areas closer to the coastline. In contrast, the UGV's operational domain is limited to land. Lastly, the UAV, being an aerial platform, is assumed to move at a constant speed without encountering any obstacles.  

Operational constraints for the aforementioned heterogeneous agents \change{are} addressed using their respective velocity maps\change{, }$V^i(\bx)$. The primary tools \change{are} the magnitude of the penalty parameter $\alpha$ in \eqref{eqn:velocity_form} and mirroring of \change{the} binary image. To begin with, it is reasonable to impose a higher penalty to the operating velocity of UUV in the proximity to land, since UUV is required to operate at a far distance from the shoreline. Thus, we set the penalty parameter to $\alpha^{uuv}=100$, while the values of $\alpha^{usv}, \alpha^{ugv}$ are set to 3. Moreover, in order to address the specific land travel limitation of the UGV, the operation domain of \change{the} UGV \change{is} obtained by the mirroring of \change{the} binary operational domain of ocean vehicles (i.e. USV, UUV), the result of which is shown in \fref{fig:reverse}. For UAVs traveling above both water and land, their domain \change{is} considered as free space without obstacles. 

The remaining parameters are the maximum operational speed $\mathcal{V}^i_{\text{max}}$ of each vehicle. In actual applications, these parameters should reflect the actual performances of agents. In this virtual test, we assume\change{ }the following scenario to demonstrate the full potential of the present approach. First, the maximum operational speed of UGV \change{is} assumed to be the slowest among all agents, considering case\change{s} where UGVs need to move as a group or encounter additional environmental restrictions (such as traffic or changes in topography). Then, we normalize\change{ the }velocities of \change{the} vehicle\change{s} using the maximum speed of UGV, and thus we write $\mathcal{V}_{\text{max}}^{ugv}=1$. The maximum speeds of USV and UUV \change{are} set to \change{the} same \change{value} $\mathcal{V}_{\text{max}}^{uuv}=\mathcal{V}_{\text{max}}^{usv}=2$, and the UAV \change{is} assumed to have the highest navigation speed\change{ }and set to $\mathcal{V}_{\text{max}}^{uav}=3$.\change{ }With the prescribed setting, we \change{apply }Algorithm~\ref{al:fmm_for_multiagent} to solve the optimization problem of rendezvous path planning. 

\subsection{Results}
\label{sec:results}

We execute\change{ }the first FMM (of the FMS) for each vehicle from arbitrarily selected initial points, as seen by the red dots in \fref{fig:uuv_time} \change{to} \fref{fig:uav_time}. The computed time grid $T^{i}(\bx)$ for each vehicle is also visualized in the same plots. One distinguished case is \fref{fig:uav_time} which shows unimpeded paths for the UAV throughout the environment. 

\begin{figure}[]
    \centering
    \includegraphics[width=0.45\linewidth]{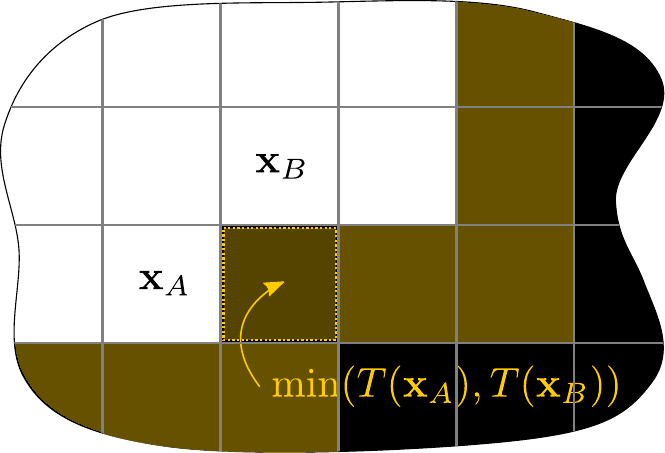}
    \caption{A\change{n illustration} that outlines procedures for extending the time grid to accommodate situations where vehicles operate in non-intersecting domains, specifically the UUV/USV and UGV.}
    \label{fig:adjacent}
\end{figure}
\begin{figure}[]
    \centering
    \includegraphics[height=0.28\textheight,trim={6cm 2.5cm 9.5cm 1.5cm},clip]{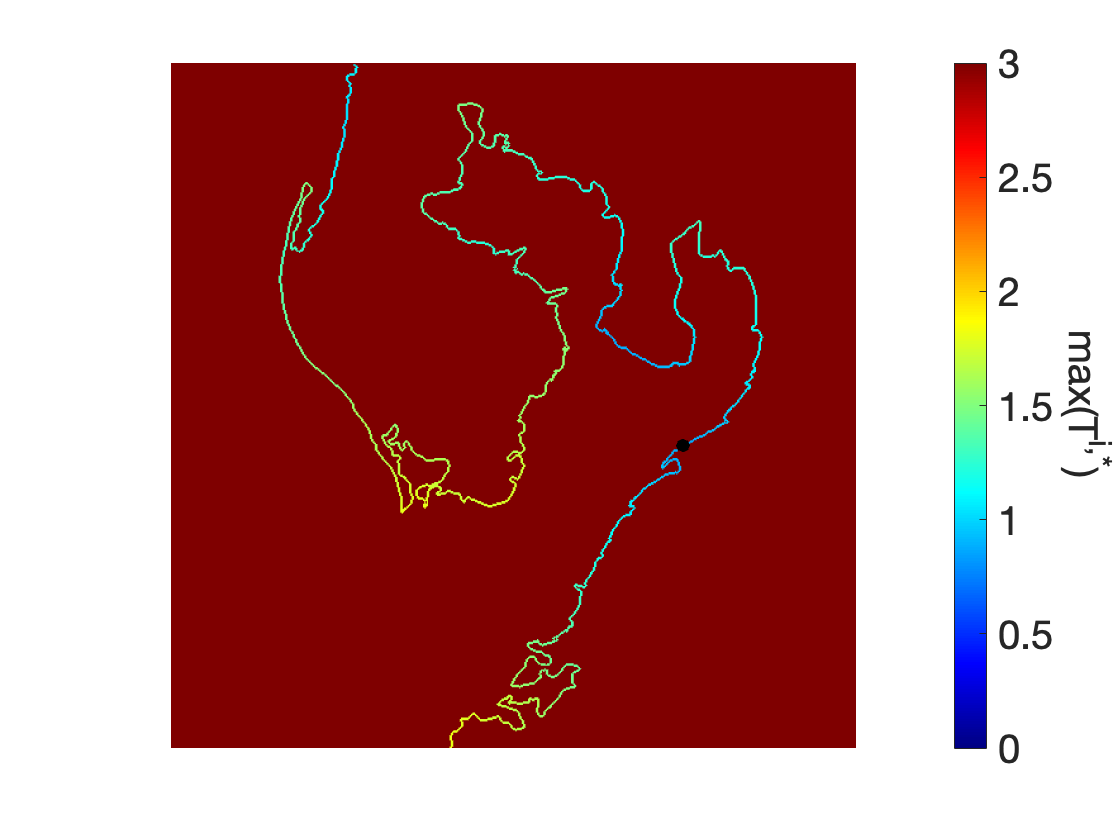}
    \includegraphics[height=0.28\textheight,trim={34cm 2.5cm 0cm 1.5cm},clip]{figures/kim16.png}
    \caption{An extended time grid defined on shoreline to compute the rendezvous point $\bx_{op}$, following the form~\eqref{eqn:optimal_point}.
   }
    \label{fig:rendezvous}
\end{figure}

\begin{figure}
    \centering
    \includegraphics[width=0.48\textwidth,trim={5cm 5cm 6cm 1.8cm},clip]{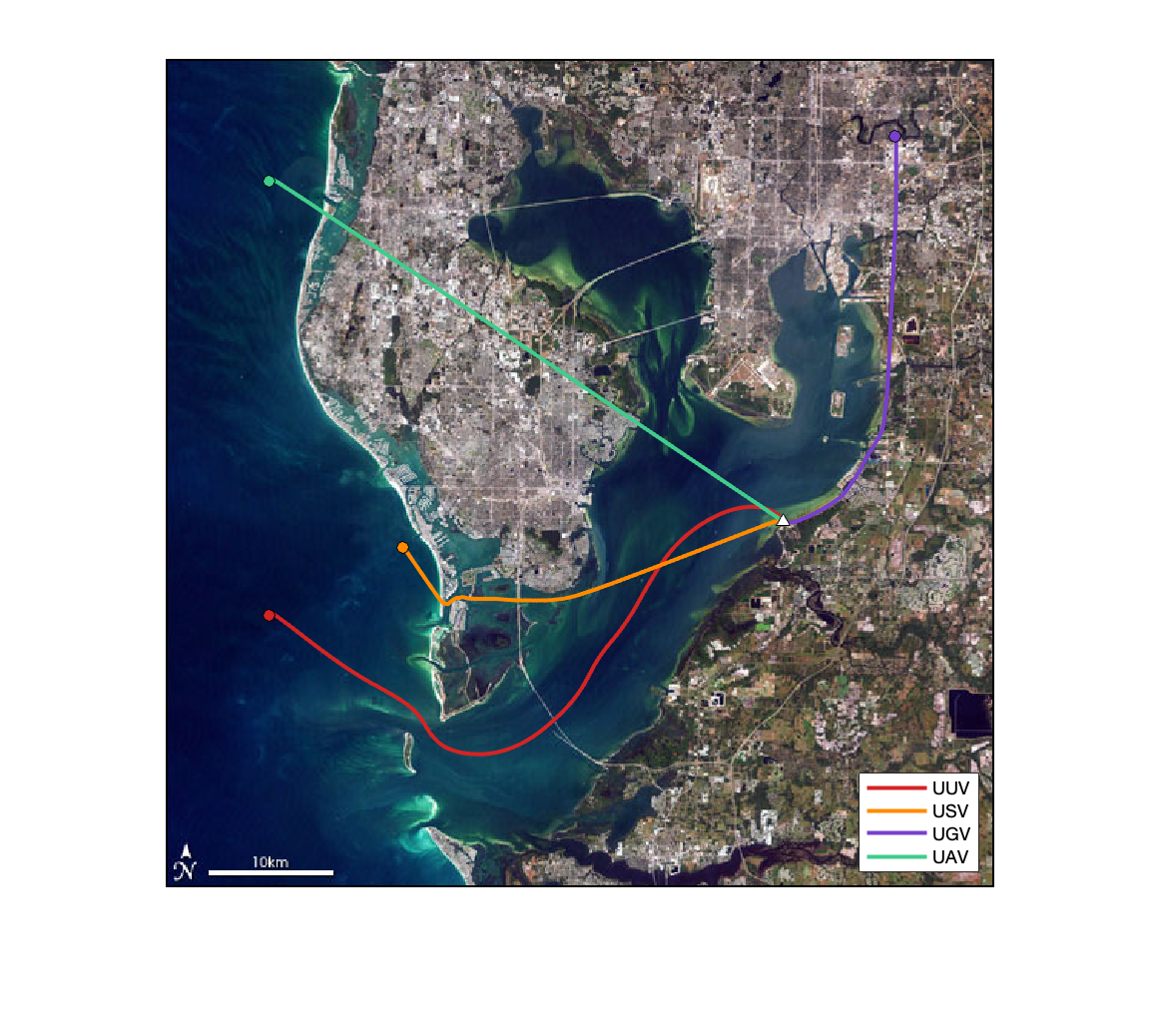}
    \caption{Path planning results plotted over the original satellite image. Circles denote starting points of vehicles, and the triangle denotes the computed optimal rendezvous point from the presented algorithm. }
    \label{fig:satellite_path}
\end{figure}

\begin{figure*}[]
    \subfloat[
          \label{fig:uuvpath}]{\includegraphics[height=0.24\textheight,trim={7cm 3cm 8.5cm 2cm},clip]{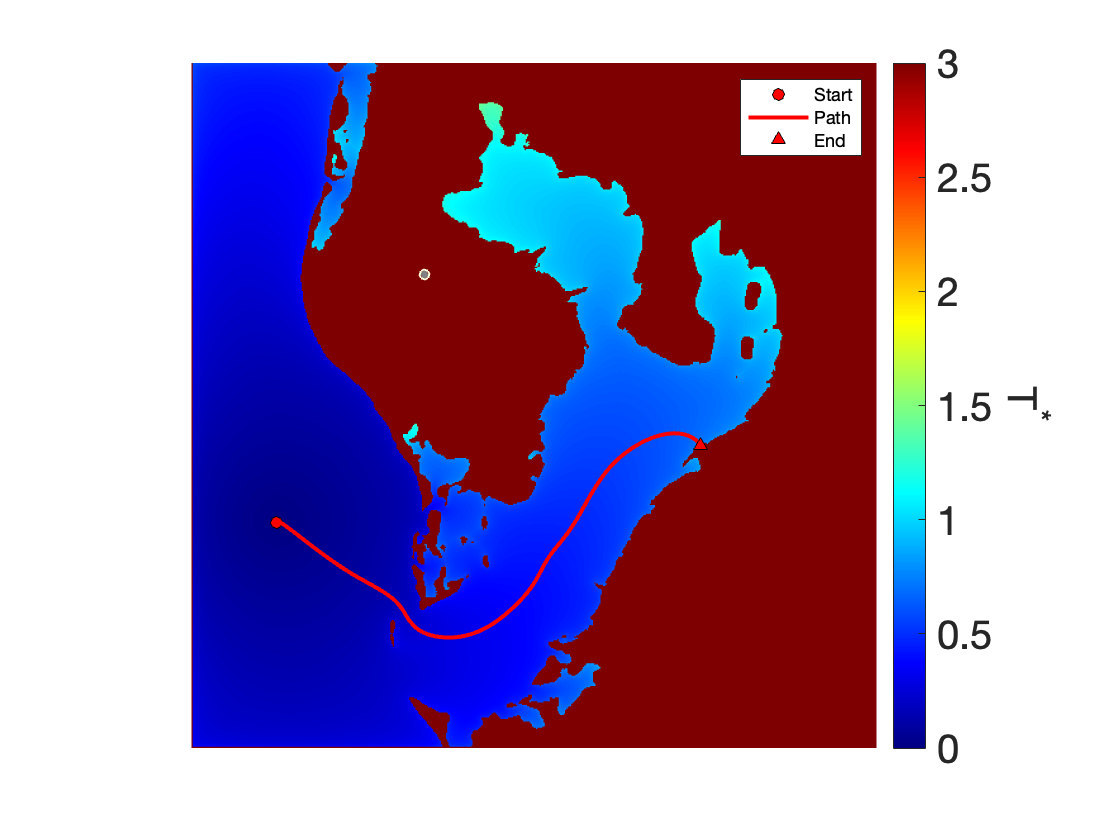}}
    \hfill
    \subfloat[
          \label{fig:usvpath}]{\includegraphics[height=0.24\textheight,trim={7cm 3cm 8.5cm 2cm},clip]{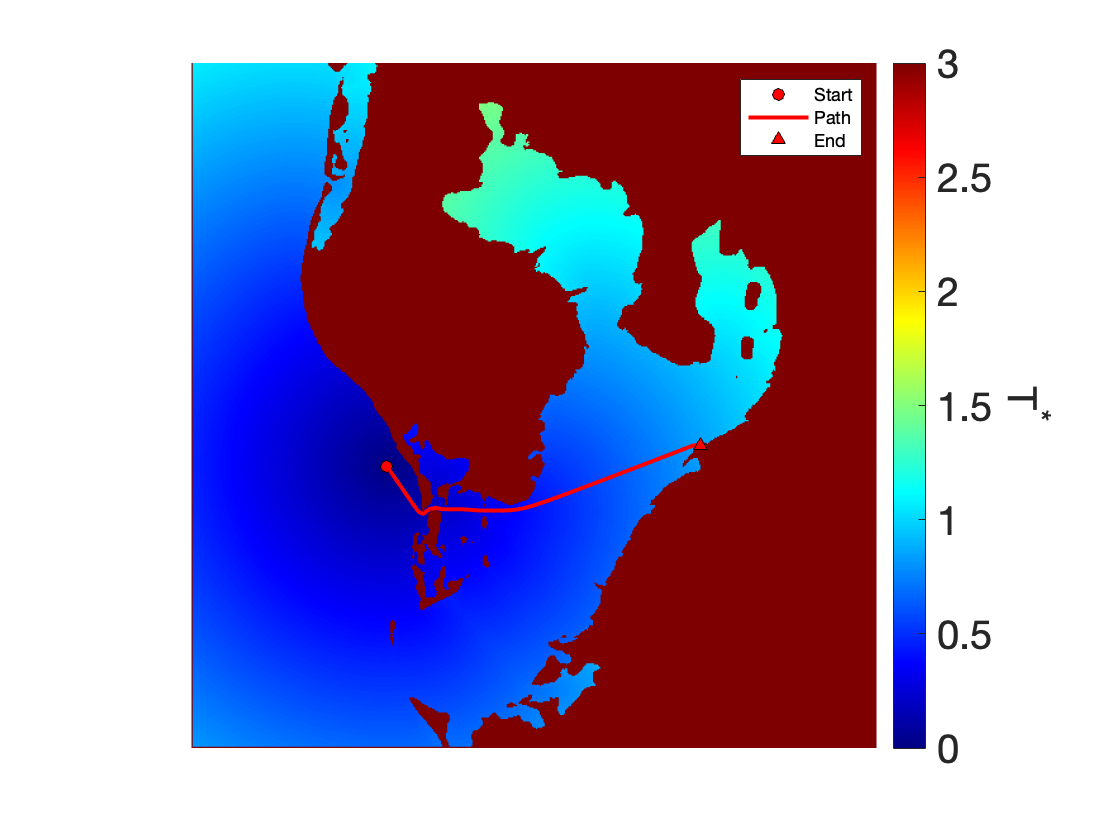}}
    \hfill
    \subfloat[
          \label{fig:ugvpath}]{\includegraphics[height=0.24\textheight,trim={7cm 3cm 8.5cm 2cm},clip]{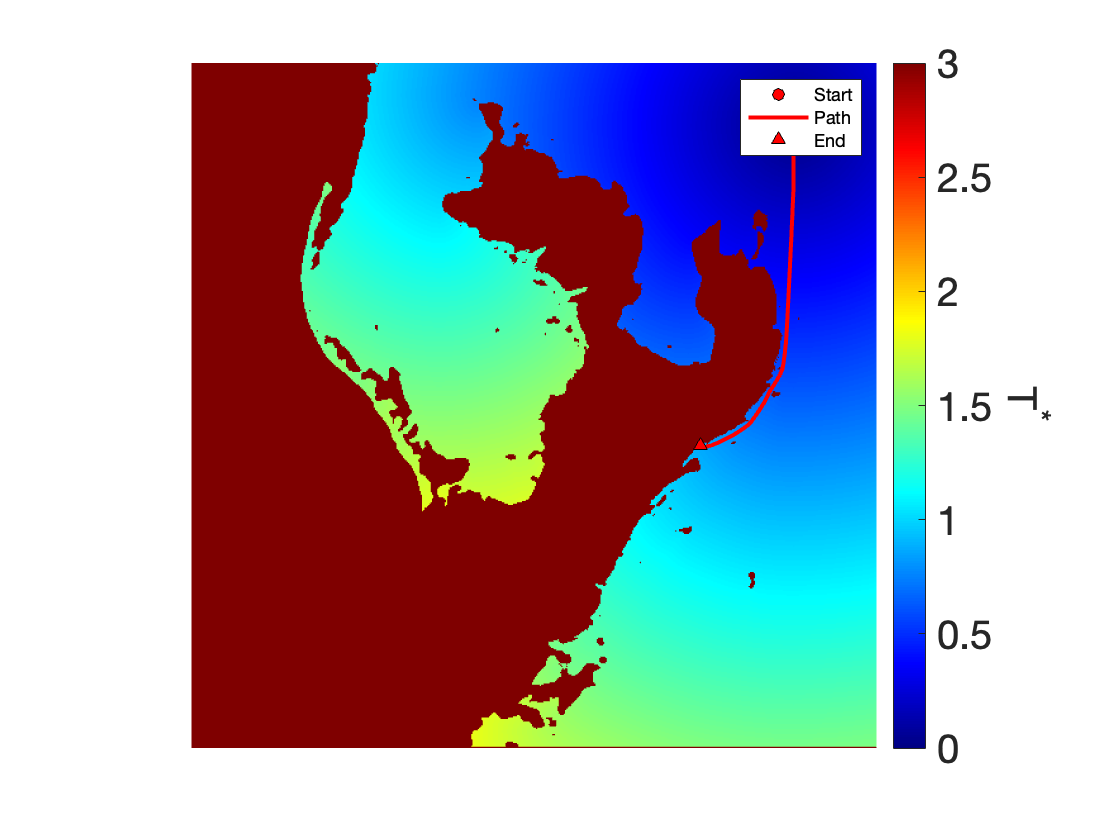}}
    \caption{The paths planned for the (a) UUV, (b) USV, (c) UGV, shown over the each time grids~\fref{fig:time_grids}.}
    \label{fig:paths}
\end{figure*}

Next, we describe an additional procedure that is not stated in Algorithm~\ref{al:fmm_for_multiagent}, but \change{is} required for the present problem, wherein the UGV operates on the complementary domain of the UUV and USV. Note that candidates for the rendezvous point \change{are} located on the shoreline, which belongs to the obstacles (i.e. $\Omega \backslash \mathcal{C}_{\text{free}}$) in the first FMM for any agent. To address \change{this} issue, we extend\change{ }the time grid $T^i(\bx)$ to the edges of obstacles. Grid cells on the edge of binary images (\fref{fig:binary} and \fref{fig:reverse}) \change{are} first detected using \texttt{MATLAB}'s ``\texttt{edge}'' function, and the $T^{i}(\bx)$ on edges \change{are} inferred from the minimum time value among the adjacent grid points. \fref{fig:adjacent} illustrates this procedure. This \change{process} results in a subset of the shoreline emerging as the candidate for the rendezvous point as seen in \fref{fig:rendezvous}. Then, in the same figure, the black dot, which represents the optimal rendezvous point where all vehicles can converge in the minimum time, is determined by the form~\eqref{eqn:optimal_point}. 

\JKadd{We also note here that we still need to compute the full arrival time map in $\mathcal{C}_{\text{free}}$. The reason is because the ultimate outcome of the present algorithm is not only to determine the rendezvous point but also to optimize the path for each agent. The latter necessitates computing arrival times for every grid point between the initial point and the rendezvous point.} 

Overall, \change{the results show} that the resulting paths in \fref{fig:satellite_path} align well with the assumed operational constraints of the member vehicles, offering realistic planning outcomes. While the paths include overlaps between the USV and UGV, path conflicts are not an issue in the test scenario\change{ because} the agents \change{operate in distinct} domains; they would not collide even if their paths overlap at the same time.

Next, the algorithm computes a path for each vehicle using the gradient descent method. \fref{fig:uuvpath} illustrates the UUV's trajectory, characterized by significant turns to remain within its operational range, attributable to the low alpha value. \change{In contrast}, \fref{fig:usvpath} exemplifies the USV's efficiency in navigating between islands. The UGV's path in \fref{fig:ugvpath} remains confined to land, adhering to its intended operational domain. Lastly, the paths for all vehicles are summarized in \fref{fig:satellite_path}, which also includes the simplest UAV's path unhindered by obstacles due to its aerial capabilities. 

\section{Discussion}
\label{sec:discussion}

In this section, we highlight the merits of our methodology in various perspective \change{and analyze the limitations}. 
\change{First, the numerical experiment presented in Section~\mbox{\ref{sec:experiment}} demonstrates how the proposed path planning method can be applied to real-world situations that demand rapid and coordinated action across various domains. For example, in disaster response scenarios, the method efficiently coordinates rescue teams composed of diverse vehicles, ensuring rapid assembly at critical locations to save lives and deliver aid. Environmental monitoring can benefit from this algorithm by synchronizing different types of sensors to converge on pollution sources or wildlife areas, facilitating comprehensive data collection. In military operations, the algorithm enhances the strategic coordination of diverse units, allowing for swift assembly and execution of missions with heightened precision and safety.} 

The authors are aware that constructing collision-free paths for agents is one of the most critical aspects of MAPF, even though our test scenario was free of such conflicts. However, even if agents share the same operational domain,\change{ }our framework \change{makes} it\change{ }easy to check whether overlapping paths will lead to collisions. This is because we have explicit information of each vehicle's arrival time at overlapping points; path conflicts occurring at different times do not result in collisions between agents. This is useful information for extending the current framework \change{to achieve} truly collision-free path planning. 

We re-emphasize that the main contribution \change{of} the present work is the formulation of \change{the} multi-agent rendezvous \change{problem} in the form of \eqref{eqn:F_example}. While Algorithm~\ref{al:fmm_for_multiagent} is straightforward under this formulation, a potential improvement can be found in the design of the velocity function~\eqref{eqn:velocity_form} to \change{consider }more detailed and realistic operational constraints of various types of vehicles. 

The main advantage of our approach is that the process is deterministic, implying that the resulting paths are guaranteed to be the globally optimized solution. 
In addition, the computational cost for determining paths of all agents is only proportional to the number of agent\change{s} $N$. 
Therefore, at least for the rendezvous task as defined in \eqref{eqn:F_example}, we claim that our approach outperforms heuristic,\change{ }stochastic, and machine learning-based methods in terms of \change{providing a} unique solution\change{ }and scalability. 

\change{
Although optimization for energy efficiency is not within the scope of this work, we expect that it can be considered, at least in a limited sense, by modeling the velocity field to incorporate physical conditions. 
For example, a drone’s most energy-efficient path can vary depending on wind conditions, while a UUV’s most energy-efficient path can vary depending on ocean currents. 
Such environmental conditions can be modeled into the operating velocity grid by adding the adverse effect of the ocean current or wind. 
However, such approach is still restricted to the shortest path corresponds to the most efficient path, since the FMM-based method is inherently grounded on a minimum-time path finding problem. Therefore, a new formulation would be necessary to determine the most energy-efficient path. Developing such a formulation and methodology require to include vehicle dynamics and control inputs as important constraints. 
This related but distinct problem formulation and solution approach will mark a significant advancement in FMM-based path planning algorithms.}

\section{Conclusion}
\label{sec:conclusion}
Recent rapid advancements in uncrewed vehicle technology have significantly improved accessibility and cost-effectiveness, leading to their widespread integration across various domains, including ground, water, and air. As systems with uncrewed vehicles become ubiquitous, the demand for sophisticated navigation methodologies that can efficiently guide their interactions also becomes paramount\change{.} In this regard, the present work \change{introduces} a new approach to path planning for multi-agent systems. Our method is rooted in the well-established framework of the FMM. The methodology presented in this paper leverages the capabilities of the FMM to efficiently optimize trajectories for heterogeneous teams of agents, augmenting their operational efficiency and collective synergy. 

To illustrate our approach, we consider\change{ }an example path planning scenario involving four different types of uncrewed vehicles navigating around the Tampa Bay area. The results of \change{the} virtual experiment \change{have} demonstrated how the path planning task of a multi-agent system can benefit from the effectiveness of the FMM-based method, which conveniently incorporates the individual operational characteristics of the heterogeneous vehicles. The computational efficiency and flexibility of our approach open the door to various directions for future work. 

\begin{itemize}
    \item The optimization function $\mathcal F$ can also be extended to incorporate various scenarios of rendezvous tasks \change{beyond} minimal time. 
    For example, we plan to include different operational costs for heterogeneous vehicles to maximize the economic efficiency of rendezvous tasks.
    \item The proposed framework can be extended to path planning in \change{the} presence of dynamic obstacles. This generalization will allow the algorithm to consider the collision between\change{ }different agents. The future work will investigate how the fast marching method can be modified in order to efficiently incorporate moving objects or other moving agents in the computation.
    \item Finally, we also envision extending our framework to different purposes of path planning for heterogeneous agents beyond rendezvous missions. This \change{extension} could involve group search optimization and assignment tasks.  
\end{itemize} 

\section*{Acknowledgment}
The authors would like to thank Blake Sanders for his help with creating Figures 13(b) and 13(c).

\printbibliography

\begin{IEEEbiography}[{\includegraphics[width=1in,height=1.25in,clip,keepaspectratio]{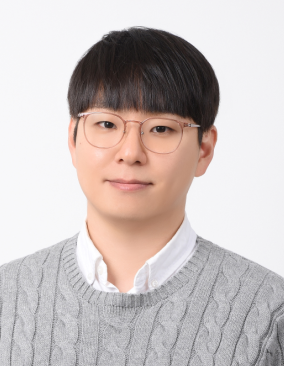}}]{Jaekwang Kim} received the B.S. degree in Naval Architecture and Ocean Engineering from Seoul National University in 2017. He received M.S. and Ph.D. degrees in Theoretical and Applied Mechanics from University of Illinois at Urbana-Champaign in 2023.  
He is now an assistant professor in the Department of Mechanical and Design Engineering at Hongik University.

\end{IEEEbiography}

\begin{IEEEbiography}[{\includegraphics[width=1in,height=1.25in,clip,keepaspectratio]{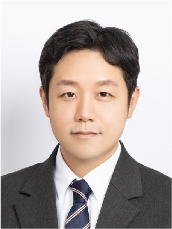}}]{Hyung-jun Park}received the B.S. degree in Naval Architecture and Ocean Engineering from Inha University in 2015. He received M.S. and Ph.D. degrees in Mechanical Engineering from Korea Advanced Institute of Science and Technology in 2021. He is now an assistant professor in the School of Mechanical and Aerospace Engineering at Sunchon National University.
\end{IEEEbiography}

\begin{IEEEbiography}
[{\includegraphics[width=1in,height=1.25in,clip,keepaspectratio]{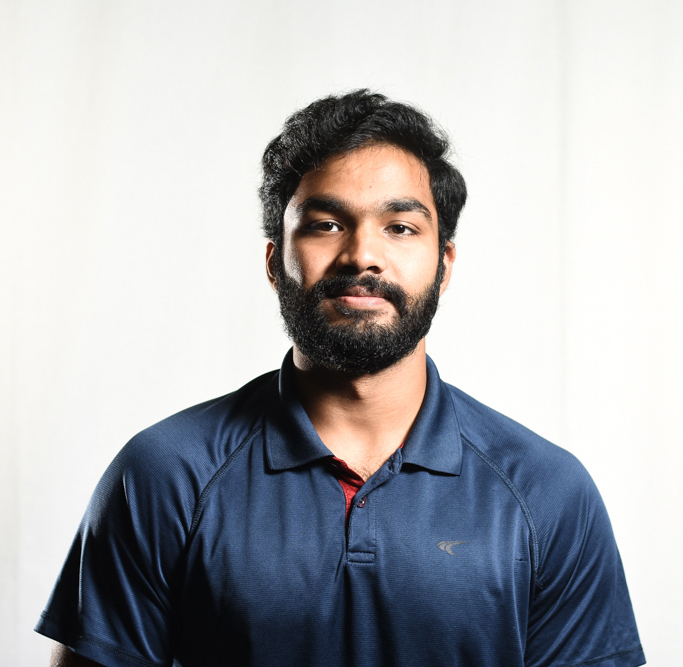}}]{Aditya Penumarti} earned a B.S. in Mechanical Engineering from North Carolina State University in 2021, followed by an M.S. in Mechanical Engineering from the University of Florida in 2023. Currently, he is a Ph.D. student in the Active Perception and Robot Intelligence Lab (APRILab) at the University of Florida under the supervision of Dr. Jane Shin.
\end{IEEEbiography}

\begin{IEEEbiography}[{\includegraphics[width=1in,height=1.25in,clip,keepaspectratio]{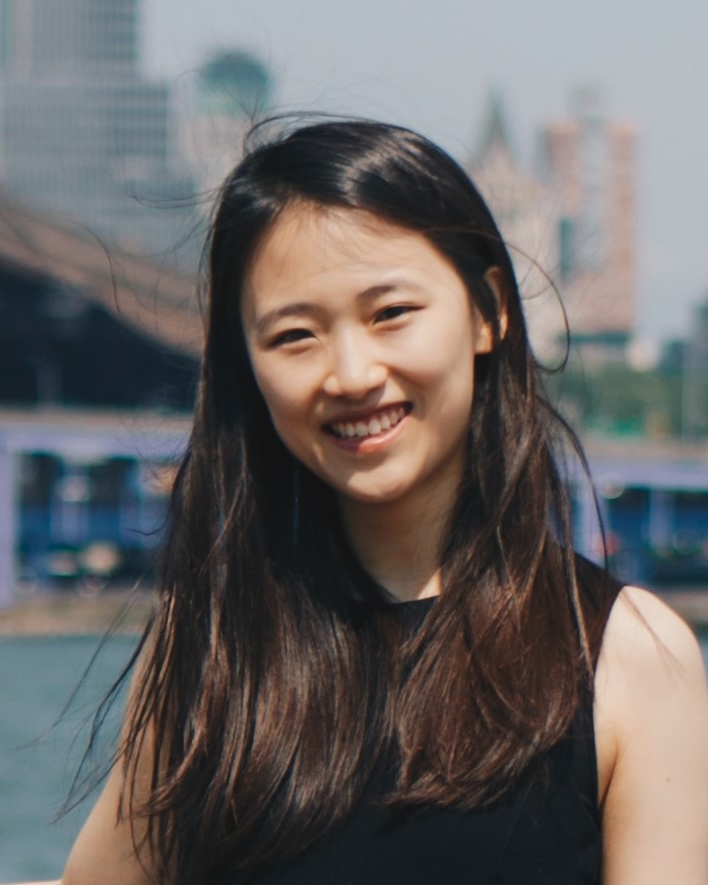}}]{``Jane'' Jaejeong Shin} received the B.S. degree in Naval Architecture and Ocean Engineering from Seoul National University in 2017. She received M.S. and Ph.D. degrees in Mechanical Engineering from Cornell University in 2021. Since 2021, she has been an Assistant Professor in the Department of Mechanical and Aerospace Engineering at the University of Florida.
\end{IEEEbiography}

\EOD

\end{document}